\documentclass[a4paper,11pt]{article}

\usepackage[T1]{fontenc} 
\usepackage[english]{babel}
\usepackage{hyphenat}
\usepackage{bm}
\usepackage{amssymb}
\usepackage{amsmath}
\usepackage{amsfonts}
\usepackage{dsfont}
\usepackage{amsbsy}
\usepackage{graphicx}

\usepackage{nccmath}

\usepackage{cancel}
\usepackage{color}
\usepackage{wrapfig}

\def\stau{$\tilde{\tau}_1$}

\def\stop{$\tilde{t}_1$ }

\def\LSP {$\chi^0_1$}

\newcommand{\slsh}[1]{\not{\hbox{\kern-2pt${#1}$}}}

\newcommand{\ba}[1]{\begin{eqnarray} \label{#1}}
\newcommand{\ea}{\end{eqnarray}}

\def\bea{\begin{eqnarray}}
\def\eea{\end{eqnarray}}
\def\bqu{\begin{quote}}
\def\equ{\end{quote}}
\newcommand{\newc}{\newcommand}
\newc{\ra}{\rightarrow}
\newc{\lra}{\leftrightarrow}

\newc{\sm}{Standard Model}
\newc{\smd}{Standard Model}
\newc{\barr}{\begin{eqnarray}}
 \newc{\earr}{\end{eqnarray}}

\def\gappeq{\mathrel{\rlap {\raise.5ex\hbox{$>$}}
{\lower.5ex\hbox{$\sim$}}}}

\def\lappeq{\mathrel{\rlap{\raise.5ex\hbox{$<$}}
{\lower.5ex\hbox{$\sim$}}}}

\begin{document}

\pagestyle{empty}

\begin{flushright}
KCL-PH-TH/2020-08, CERN-TH-2020-023 \\
UHU-CFMC/2020-07\\

\end{flushright}

\vspace*{0.5cm}
\begin{center}
{\Large {\bf 
Confronting Grand Unification with Lepton Flavour Violation,  \\
\vspace*{0.2cm}
Dark Matter and LHC Data}} \\
\vspace*{1cm}
{\bf J.~Ellis$^{1}$, M.E. G{\'o}mez$^2$, S.~Lola$^{3}$, R.~Ruiz~de~Austri$^4$, Q.~Shafi$^5$}\\
\vspace*{0.5cm}
$^1$ Theoretical Particle Physics and Cosmology Group, Department of Physics,
\\ King's College London, Strand, London WC2R 2LS, UK;\\
National Institute of Chemical Physics \& Biophysics, R\"avala 10, 10143 Tallinn, Estonia;\\
Theoretical Physics Department, CERN, CH-1211 Geneva 23, Switzerland \\
\vspace*{0.2cm}
$^2$ Departamento de Ciencias Integradas y Centro de Estudios Avanzados en F\'{i}sica Matem\'aticas y Computaci\'on, Campus El Carmen, Universidad de Huelva, 21071 Huelva, Spain \\
\vspace*{0.2cm}
$^3$ Department of Physics, University of Patras, 26500 Patras, Greece \\
\vspace*{0.2cm}
$^4$ Instituto de F\'isica Corpuscular, IFIC-UV/CSIC, Valencia, Spain \\
\vspace*{0.2cm}
$^5$ Bartol Research Institute, Department of Physics and Astronomy, University of Delaware, \\ Newark, DE 19716, USA \\

\vspace*{0.3cm}
%
%
%
%
%
%
\vspace*{1.2cm}
{\bf ABSTRACT} \\ 
\end{center}
We explore possible signatures for charged lepton flavour violation (LFV), sparticle discovery at the LHC and dark matter (DM) searches in grand unified theories (GUTs) based on SU(5), flipped SU(5) (FSU(5)) and SU(4)$_c \times $SU(2)$_L \times $SU(2)$_R$ (4-2-2).
We assume that soft supersymmetry-breaking terms
preserve the group symmetry at some high input scale, and focus on the non-universal effects on different matter representations generated by gauge interactions at lower scales, as well as the charged LFV induced in Type-1 see-saw models of neutrino masses. We identify the different mechanisms that control the relic DM density in the various GUT models, and contrast their LFV and LHC signatures.
The SU(5) and 4-2-2 models offer good detection prospects both at the LHC and in LFV searches, though with different LSP compositions, and the SU(5) and FSU(5) models offer LFV within the current reach.
The 4-2-2 model allows chargino and gluino coannihilations with neutralinos, and the former offer good detection prospects for both the LHC and LFV, while gluino coannihilations lead to lower LFV rates.
Our results indicate that LFV is a powerful tool that complements LHC and DM searches, providing significant insights into the sparticle spectra and neutrino mass parameters in different models.


\newpage
\noindent

\setcounter{page}{1}
\pagestyle{plain}

\section{Introduction}
\label{sec:1}

Experimental and theoretical considerations both require extending
the Standard Model (SM) of particle physics, which can neither
accommodate massive neutrinos, nor explain the observed
baryon asymmetry of the universe, nor provide dark matter \cite{WMAP1,WMAP2,Ade:2013zuv,Ade:2015xua}.
Nevertheless, the data from the LHC~\cite{higgs1,higgs2,CMSdat,ATLASdat}  and dark matter
searches~\cite{LUX,XENON1T,Cui:2017nnn,XENON1T_new,PICO} 
have not yet yielded any positive signature of physics beyond the SM. On the contrary, severe constraints have been derived for
the simplest extensions of the SM that address these issues,
including the most simplified versions of supersymmetric unified theories.
However, supersymmetry (SUSY) continues to have strong theoretical attraction and,  
among other features, provides a natural candidate for dark matter (DM) \cite{Ellis:1983ew} and facilitates the
construction of grand unified theories (GUTs)~\cite{GUT}.
It is therefore premature to exclude SUSY before 
studying in more detail non-simplified models
that have not yet been explored.

In doing so, flavour physics inevitably plays a crucial
role, since it also provides severe bounds on
extensions of the SM that would have resulted in exotic manifestations of
flavour violation that have not yet been observed.
In particular, supersymmetric theories have several 
possible sources of lepton flavour violation (LFV), 
which would yield unacceptably large effects 
unless off-diagonal entries in the sfermion mass matrices 
were small at some high scale.
Even in this case, however,  quantum corrections during the 
running from high scales to low energies would modify this
simple structure. This effect is particularly significant
in see-saw models for neutrino masses, where
the Dirac neutrino Yukawa couplings cannot be diagonalised 
simultaneously with the charged (s)lepton Yukawa couplings \cite{Borzumati:1986qx}. In this case, 
the large mixing of neutrino families required by the data also
implies that charged LFV may occur 
at enhanced rates for sufficiently small
soft supersymmetry-breaking masses. This can occur in
rare decays and conversions (e.g., $\mu \rightarrow e \gamma$ and 
$\tau \rightarrow \mu \gamma$,
$\mu \rightarrow 3 e$,  $\tau \rightarrow e \gamma$ and
$\mu - e$ conversion \cite{MEG2013,PDG}), ~\footnote{For pioneering studies of $\mu \rightarrow e \gamma$ in supersymmetric models, see~\cite{Lee}.} but also in other processes
such as sparticle production at the LHC
\cite{LHC1,LHCh,LHC2,Bartl,CEGLR2,AbadaLHC,EstevesLHC,HirschLHC}
and slepton
pair production at a Linear Collider (LC)~\cite{sleptonscemu,LFV-LCa,LFV-LCb,LC2,LFV-LC2,CannoniLC,CEGL-LC,Abada}, particularly in the decays of the second-lightest
neutralino. 

In this paper we study the possibilities for LFV, LHC and dark matter
searches in models with
various grand unification groups.
We pay particular attention to comparisons between their respective
signatures, and to ways to differentiate between the various schemes
in present and future searches. Since the SO(10) GUT model is now severely
constrained by the data, 
we focus on GUTs based on SU(5), flipped SU(5) (FSU(5)) and
SU(4)$_c \times $SU(2)$_L \times $SU(2)$_R$ (4-2-2) \cite{PS,PS-lr,PS2}.
Similarly to previous works within various GUT models 
\cite{EGLR,Okada:2013ija,
Kowalska:2015zja,Kowalska:2014hza,Ellis:2016tjc, Gomez:2018efz,Gomez:2018zzw},
we assume that at the unification scale the soft SUSY-breaking terms
preserve the group symmetry, and focus on the non-universal effects 
on different matter representations due to the gauge structures of the groups. We also include the charged LFV induced in these models via Type-1 see-saw models of neutrino masses with right-handed neutrinos at some high intermediate mass scale. 

Different mechanisms that control the relic DM density in the various GUT models, lead to contrasting LFV and LHC signatures. Although the results are sensitive to the scale of the right-handed neutrinos (with larger scales being linked to larger couplings and thus larger flavour violation due to quantum corrections), for similar see-saw parameters, detailed comparisons between different unification schemes can be made. In all cases, coannihilations result to higher LFV rates. We find that
the SU(5) and 4-2-2 models have different LSP compositions, but both offer good prospects for detection at both the LHC and in LFV searches. 
The FSU(5) model also offers LFV within the current reach, for instance in stop and stau coannihilation scenarios.
The 4-2-2 model admits novel DM mechanisms such as chargino and gluino coannihilations with neutralinos. The former offers good detection prospects for both the LHC and LFV, whereas gluino coannihilations lead to lower LFV rates, and Higgsino DM models do not predict detectable LFV. 
In addition, we derive specific correlations between the respective sparticle spectra, providing
further input on the experimental signatures that can be expected in each scheme, commenting on
the prospects for direct detection of DM as well as LHC and LFV searches.

In Section 2 we summarise the basic non-universal features of the GUTs we study
that are relevant for our discussion. In Section 3 we discuss different
mechanisms for determining the relic density of dark matter (DM) in the presence
of non-universal SUSY-breaking mass terms.
In Section 4 we discuss lepton flavour mixing effects in the presence
of see-saw neutrinos. In Section 5 we look at the branching ratios for
rare LFV decays in different DM scenarios, also taking LHC and direct
DM searches into account. Finally, our
main results and future detection prospects 
are summarised in Section 6.

\section{Non-universal soft supersymmetry breaking in GUT models}
\label{sec:3}

In our analysis, we assume that SUSY breaking 
occurs in a hidden sector at some scale $M_X > M_{GUT}$, via a mechanism that 
generates flavour-blind soft terms in the visible sector. Between the scales 
$M_X$ and $M_{GUT}$, although the theory still preserves the GUT symmetry,
quantum corrections
may induce non-universalities for soft terms that belong to
different GUT representations, while particles that belong to 
the same representation have common soft masses.

The soft SUSY-breaking scalar terms for the fields in an irreducible representation $r$ of the 
unification group are parametrised as multiples of a common scale $m_0$: 
\begin{equation}
m_{r}=x_r \, m_{0}, 
\end{equation}
while the trilinear terms are defined as: 
\begin{equation}
A_r = Y_r \,  A_0,  \;\;\; A_0=a_0 \, m_0 \, ,
\end{equation}
where $Y_r$ is the Yukawa coupling associated with the
representation  $r$. We use the standard parametrization with $a_0$
a dimensionless factor, which we assume to be representation-independent. 
Since the two Higgs fields of the MSSM arise from different SU(5)
representations, they have
in general   different soft masses. The situation in the different GUT groups is then as follows:

$\bullet $ SU(5): 
In this case, the multiplet assignments are as follows:
\begin{equation}
(Q,u^{c},e^{c})_{i} \; \in \mathbf{10}_i \, , \; (L,d^{c})_{i} \; \in \; \mathbf{\overline{5}}_i \, , \; \nu^c_{i} \; \in \; \mathbf{1}_i \, .
\end{equation}
We assume that the soft terms are the same for all the members of the
same representation at the GUT scale, but may be different for the $\mathbf{10}$ and 
$\mathbf{\overline{5}}$ representations. Here we  use as reference the common soft
SUSY-breaking  masses for the 
fields of the ${\bf 10}$, denoted by $m_{10}$. The masses for the other
representations  are then defined as:
\begin{flalign}
 m_{10}=m_0, \;\;\; m_5=x_5\cdot m_{10},  \;\;\; m_{H_u}=x_u \cdot m_{10},  \;\;\; m_{H_d}=x_d \cdot m_{10}  \, ,
\end{flalign}
and the $A$ terms are specified via a common mass scale:
\begin{equation}
A_{10,5}=a_0 \, m_{0} \, .
\end{equation}

$\bullet $ Flipped SU(5) (FSU(5)): 
Since the particle assignments are now different, namely:
\begin{equation}
(Q,d^{c},\nu^{c})_{i} \; \in \mathbf{10}_i \, , \; (L,u^{c})_{i} \; \in \; \mathbf{\overline{5}}_i \, , \; e^c_{i} \; \in \; \mathbf{1}_i \, ,
\end{equation}
and the parametrisation changes to
\begin{equation}
m_{10}=m_0, \;\;\; m_5=x_5\cdot m_{10}  \;\;\; m_R=x_R\cdot m_{10}  \;\;\; 
m_{H_u}=x_u \cdot m_{10}  \;\;\; m_{H_d}=x_d \cdot m_{10}  \, ,
\end{equation}
where $x_R$ refers to the SU(2)-singlet fields.
As previously, the $A$ terms are specified as universal: $A_0=a_0\cdot
m_0$. 

$\bullet $ 4-2-2 symmetry: 
In this case, a significant modification arises already in the correlation of
gaugino masses, since the embedding of the hypercharge generator in
the 4-2-2 group implies:
\begin{equation}
M_1=\frac{3}{5} M_2 + \frac{2} {5} M_3 \, ,
\label{eq:M1}
\end{equation}
yielding gluino coannihilations that
are absent in models based on other groups \cite{EGLR}.
Sfermions are accommodated in 16-dimensional spinor
representations, with their common soft mass parameter being $m_{16}$. 
The electroweak MSSM doublets
lie in the 10-dimensional representation with D-term contributions
that split their soft masses: 
$m^2_{H_{u,d}} = m^2_{10} \pm 2M_D^2$. In our notation: 
\begin{equation}
x_u=\frac{m_{H_{u}}}{m_{16}}, \;\;\;\;\;\;\;\;\;\;
x_d=\frac{m_{H_{d}}}{m_{16}}, \;\;\;\;\;\;\;\;\;\;
\end{equation}
with  $x_u<x_d$.
In the left-right asymmetric 4-2-2 model, a new parameter is introduced:
\begin{equation}
x_{\,_{LR}}=\frac{m_L}{m_R}, \;\;\;\;\;\;\;\;\;\;
\end{equation}
where $m_L$ is the mass of the left-handed sfermions (that preserve
the definition of  $m_{16}=m_0$), and $m_R$ is the mass of the corresponding 
right-handed ones.

\section{Relic density mechanisms and GUT mass relations}
\label{sec:4}

We assume the following relic density
constraint~\cite{Ade:2015xua}:
\begin{equation}
\Omega_\chi h^2 \; = \; 0.1186\pm 0.0031 \, ,
\label{DMdensity}
\end{equation}
with a (fixed) theoretical uncertainty of 
$\tau= 0.012$, (following Refs.~\cite{Roszkowski:2014wqa})
to account for numerical uncertainties in the relic
density calculation. This narrow range on the relic
density imposes a strong constraint on the DM
candidate and the mechanisms that determine 
its density.

It is well known that
particular mass relations must be present in the 
supersymmetric spectrum if the required 
amount of relic dark matter is provided by neutralinos. In addition to mass relations, we use the neutralino composition
to classify the relevant points of the supersymmetric parameter space.
The higgsino fraction of the lightest neutralino mass eigenstate is characterized by
the quantity
\begin{flalign}
h_f \; \equiv \; |N_{13}|^2 + |N_{14}|^2 \, ,
\end{flalign}
where the $N_{ij}$ are the elements of the unitary 
mixing matrix that correspond to the higgsino mass states.
Thus, we classify the points that pass the relic density
constraint discussed above according 
to the following criteria:\\

\noindent
{\bf Higgsino DM:}
\begin{flalign}
h_f >0.1, \;\;|m_A-2 m_\chi| > 0.1 \, m_\chi.
\label{criterio_higgsino}
\end{flalign}
The first condition in (\ref{criterio_higgsino}) ensures that
the lightest neutralino is  higgsino-like and, as we discuss later, 
the lightest chargino $\chi^\pm_1$ is almost degenerate in mass with \LSP.
The couplings to the SM gauge bosons are not suppressed and \LSP\ pairs have large 
cross sections for annihilation into $W^+ W^-$ and $ZZ$ pairs, which may reproduce the observed value
of the relic DM abundance. Clearly, coannihilation channels involving $\chi^\pm_1$ and 
$\chi^0_2$ also contribute. The second condition in (\ref{criterio_higgsino}) implies that
the DM density is not controlled by rapid annihilation through $s$-channel resonances.\\
~~\\
{\bf $A/H$ resonances:}
\begin{equation}
|m_A-2 m_\chi|\leq 0.1 \, m_\chi.
\label{criterio_res}
\end{equation}
This condition ensures that the correct value
of the relic DM abundance is achieved thanks to $s$-channel annihilation,
enhanced by the resonant heavy neutral Higgs ($A$ and $H$) propagators. The thermal average $\langle \sigma_{ann}v\rangle$
spreads out the  peak in the cross section, so that neutralino masses for 
which $2m_{\chi} \simeq m_A$ is not exactly realized can also experience resonant annihilations.\\
~~\\
{\bf $\tilde{\tau}-$\LSP \,  coannihilations:}
\begin{flalign}
h_f <0.1,\;\;(m_{\tilde{\tau}_1}-m_\chi)\leq 0.1 \, m_\chi
\label{criterio_RR}
\end{flalign}
The first condition in (\ref{criterio_RR}) ensures that the neutralino is bino-like, 
in which case annihilation into leptons through $t$-channel slepton exchange 
is suppressed, and when the second condition in (\ref{criterio_RR}) is satisfied
coannihilations involving the nearly-degenerate \stau \ enhance the thermal-averaged effective cross section.\\
~~\\
{\bf $\tilde{\tau}-\tilde{\nu}_\tau-$ \LSP \, coannihilations:}
\begin{flalign}
h_f <0.1,\;\;(m_{\tilde
{\tau}_1}-m_\chi)\leq 0.1 \, m_\chi,\;\;(m_{\tilde{\nu}_\tau}-m_\chi) \leq  0.1 \, m_\chi.
\label{criterio_LL}
\end{flalign}
This is similar to the previous case, but also the $\nu_{\tilde \tau}$ is nearly degenerate in mass with the 
\stau.\\
~~\\
{\bf $\tilde{t_1} - $ \LSP \, coannihilations:}
\begin{flalign}
h_f <0.15,\;\;(m_{\tilde{t}_1}-m_\chi)\leq 0.1 \, m_\chi.
\label{criterio_stop}
\end{flalign}
In this case the \stop \ is light and nearly degenerate with the bino-like neutralino.
These coannihilations are present in the flipped SU(5) model and in
the 4-2-2 model, but not SU(5).\\

In previous work, we had found that the 4-2-2 model may be
distinguished clearly from the other GUT groups, due to  
the appearance of three additional modes of coannihilation that are not present in
  other groups: \\

\noindent
{\bf $\tilde{\chi}^+-$\LSP \, coannihilations:}
\begin{flalign}
h_f <0.1,\;\;(m_{\tilde{\chi}^+}-m_\chi)\leq 0.1 \, m_\chi.
\label{criterio_char}
\end{flalign}
In this case the Higgsino component in the \ LSP \ is small,
but the lightest chargino  is light  and nearly degenerate with the bino-like neutralino.\\
~~\\
{\bf $\tilde g -$\LSP \ coannihilations:}
\begin{flalign}
h_f <0.1,\;\;(m_{\tilde{g}}-m_\chi)\leq 0.1 \, m_\chi,
\label{criterio_glu}
\end{flalign}
In this case the gluino can be relatively light and nearly degenerate with
the bino-like neutralino. \\
~~\\
{\bf $\tilde{b}-$\LSP \, coannihilations:}
\begin{flalign}
h_f <0.1,\;\;(m_{\tilde{b}}-m_\chi)\leq 0.1 \, m_\chi,
\label{criterio_bot}
\end{flalign}
in this case, in the presence of the LR asymmetry, the $\tilde{b}$  can be light  and
nearly degenerate with a bino-like neutralino \cite{sbottom-Co}.

\section{Lepton-flavour mixing effects and see-saw neutrino masses}
\label{NR-CMSSMI}
In what follows, we supplement the previous 
framework with a see-saw mechanism so as to incorporate neutrino
masses \cite{Calibbi:2017uvl, Vicente:2015cka}. We consider a high-scale see-saw 
mechanism in which, in order to obtain order 0.1 eV masses for
the neutrinos, this scale should be around $10^{14}$GeV (assuming
electroweak-scale Dirac neutrino masses). Such a mechanism can be realized by
extending the MSSM with renormalizable interactions in three
scenarios: type I \cite{seesaw:I} that requires singlet RH neutrinos,
type II~\cite{Lazarides:1980nt, Mohapatra:1980yp} that requires scalar
$SU(2)_L$ triplets and type III \cite{Foot:1988aq} that requires
fermionic $SU(2)_L$ triplets. Here we focus on the type-I see-saw, in which the additional  singlet RH neutrino fields do not affect the running of the gauge couplings and therefore fit well in our unification schemes. Some examples of  the phenomenology of type II and type III models can be found in Refs~\cite{Esteves:2010si, Esteves:2011gk}.

    We use  the following superpotential:

\begin{eqnarray}
\label{superpotentialSeesaw1}
W&=&W_{\rm MSSM}+ Y_{\nu}^{ij}\epsilon_{\alpha \beta} H_2^{\alpha} N_i^c L_j^{\beta}
+ \frac{1}{2} M_{N}^{ij} N_i^c N_j^c,
\end{eqnarray}
\noindent
where $W_{\rm MSSM}$ is the MSSM superpotential and the $N_i^c$
are additional superfields that contain the three singlet (right-handed) neutrinos,
$\nu_{Ri}$, and 
their scalar partners, $\tilde \nu_{Ri}$, and $M_N^{ij}$ denotes the $3\times3$ Majorana mass matrix for the heavy right-handed neutrinos.
The full set of soft SUSY-breaking terms is given by
\begin{eqnarray}
\label{softbreakingSeesaw1}
-L_{\rm soft,SI} &=& - L_{\rm soft}
+(m_{\tilde \nu}^2)^i_j {\tilde \nu}_{Ri}^* {\tilde \nu}_{R}^j
+ (\frac{1}{2}B_{\nu}^{ij} M_{N}^{ij} {\tilde \nu}_{Ri}^* {\tilde \nu}_{Rj}^*
+A_{\nu}^{ij}h_2 {\tilde \nu}_{Ri}^* {\tilde l}_{Lj}+ {\rm h.c.})~,
\end{eqnarray}
where $L_{\rm soft}$ contains the MSSM soft SUSY-breaking masses, 
and $(m_{\tilde \nu}^2)^i_j$,  $A_{\nu}^{ij}$ and $B_{\nu}^{ij}$
are the new soft SUSY-breaking parameters in the see-saw sector.

The see-saw mechanism yields three heavy neutral mass eigenstates 
that are mainly right-handed and
decouple at a high energy scale, with masses that we denote as $M_N$. Below this scale, the effective theory
contains the MSSM plus a higher-dimensional operator that provides masses for the light neutrinos,
which are mainly left-handed:

\begin{equation}
W=W_{\rm MSSM}+ \frac{1}{2}(Y_{\nu} L  H_2)^{T}  M_{N}^{-1} (Y_{\nu} L  H_2).
\end{equation}
As the right-handed neutrinos decouple at their respective mass scales, at low energy we have the same particle content and mass matrices as in the MSSM. This framework 
naturally accommodates neutrino oscillations that are consistent with
experimental data~\cite{Neutrino-Osc}. At the electroweak scale an
effective Majorana mass matrix for light neutrinos, 
\begin{equation}
m_{\rm eff}=-\frac{1}{2}v_u^2 Y_{\nu}\cdot M_{N}^{-1}\cdot Y^{ T}_{\nu}, 
\label{meff}
\end{equation}
arises from the Dirac neutrino Yukawa coupling matrix $Y_{\nu}$ (with entries that can be assumed to be of
the same order of magnitude as the charged-lepton and quark Yukawa couplings), and the heavy Majorana
masses $M_N$.  

We observe from (\ref{superpotentialSeesaw1}) that we can rotate the fields $L_i$ and $N_i^c$ in
such a way that the matrices of the lepton Yukawa couplings,
$Y_l^{ij}$, and of the right-handed neutrinos, $M_N^{ij}$, become 
diagonal. However, in this basis,
the neutrino Yukawa couplings $Y_{\nu}^{ij}$ are not in general diagonal,
giving rise to lepton-flavour-violating (LFV) effects
~\cite{Cannoni:2013gq,gllv,Mismatch,Antusch,EGL,casas-ibarra,Gomez:2014uha}.
It is important to note here that lepton-flavour
conservation is not a consequence of the SM gauge symmetry, even in the absence
of the right-handed neutrinos. 
Consequently, slepton mass terms can violate
lepton-flavour conservation in a manner consistent with the gauge
symmetry.  Indeed, the scale of LFV can be identified with the
EW scale, much lower than the right-handed neutrino scale, $M_R$, which we assume to be common,
for simplicity. In
the basis where the charged-lepton Yukawa matrix $Y_{\ell}$ is diagonal, the soft
slepton mass matrix acquires corrections that contain off-diagonal
contributions from the RGE running from $M_{\rm GUT}$ down to $M_R$,
which are of the following form in the leading-log approximation~\cite{LFVhisano}: 
\begin{eqnarray}
(m_{\tilde L}^2)_{ij} &\sim & \frac 1{16\pi^2} (6m^2_0 + 2A^2_0)
\left({Y_{\nu}}^{\dagger} Y_{\nu}\right)_{ij}  
\log \left( \frac{M_{\rm GUT}}{M_R} \right) \, , \nonumber\\
(m_{\tilde e}^2)_{ij} &\sim & 0  \, , \nonumber\\
(A_l)_{ij} &\sim & \frac 3{8\pi^2} {A_0 Y_{l}}_i
\left({Y_{\nu}}^{\dagger} Y_{\nu}\right)_{ij}  
\log \left( \frac{M_{\rm GUT}}{M_R} \right) \, .
\label{offdiagonal}
\end{eqnarray}
Below $M_R$, the off-diagonal contributions remain almost unchanged.
Their magnitude
depends on the structure of $Y_\nu$ at $M_R$,
in a basis where $Y_l$ and $M_N$ are diagonal.  
Using the approach of \cite{casas-ibarra,Gomez:2015ila} a generic form for $Y_\nu$
that contains all neutrino experimental information can be obtained: 

\begin{equation}
Y_\nu = \frac{\sqrt{2}} {v_u} \sqrt{M_N^\delta} O_R  \sqrt{m_\nu^\delta} U^\dagger~,
\label{eq:casas} 
\end{equation}
where $O_R$ is a general orthogonal matrix and  $M_N^\delta$ and  $m_\nu^\delta$ denote the
diagonalized heavy and light Majorana neutrino mass matrices, respectively. In this basis the matrix~$U$ can be
identified with the Pontecorvo-Maki-Nakagawa-Sakata (PMNS) matrix, $U_{\rm PMNS}$: 
\begin{equation}
m_\nu^\delta=U^T m_{\rm eff} U~.
\end{equation}

Assuming that the observed neutrino oscillations can be attributed to
hierarchical neutrino masses, we have $m_\nu^\delta=Diag(1.1\cdot
10^{-3}, 8  \cdot 10^{-3}, 5 \cdot 10^{-2})$~eV. For $Y_\nu$ couplings of
order one, the RH neutrinos can take values as large as $10^{14}$
GeV. The LFV BR's decrease with the RH neutrino scale.
However, the matrix $O_R$ is associated with the flavor structure of the RH neutrino mass matrix, which must be nontrivial so as to provide a scenario for  baryogenesis through leptogenesis, which typically requires masses of order $10^8-10^9$~GeV \cite{Buchmuller:2004nz,Fong:2013wr}. It may also induce  cancellations in the LFV BRs  that may allow RH neutrino masses above the $10^{14}$~GeV scale while respecting the current constraints.
 To illustrate
this point, we consider real see-saw parameters and parametrize the
matrix $O_R$ with three real angles $\theta_1$, $\theta_2$,
$\theta_3$, following the notation of Ref. \cite{casas-ibarra} : 
\begin{equation}
O_R=R_{12}(\hat{\theta}_3)\cdot R_{13}(\hat{\theta}_2)\cdot R_{23}(\hat{\theta}_1),
\end{equation}
where

\begin{equation}
R_{23}=\begin{pmatrix} 1&0&0 \\
				0&\hat{c}_1& \hat{s}_1\\
				0&\hat{s}_1& \hat{c}_1		
\end{pmatrix};
 R_{13}=\begin{pmatrix} \hat{c}_2&0& \hat{s}_2\\
				0&1&0 \\
				\hat{s}_2&0& \hat{c}_2	
\end{pmatrix} ;
R_{12}=\begin{pmatrix} \hat{c}_3& \hat{s}_3&0\\
				\hat{s}_3& \hat{c}_3&0\\
				0&0&1	
\end{pmatrix} .
\label{eq:or}
\end{equation}

\noindent We denote  $\sin(\hat{\theta}_i)$ and $\cos(\hat{\theta}_i)$ as  $\hat{s}_i$ and  $\hat{c}_i$ respectively $(i=1,2,3)$. 

For the matrix $U$, we  consider as an illustrative example the Harrison, Perkins, and Scott (HPS) mixing matrix \cite{Harrison:2002er}: 
\begin{equation}
U=\begin{pmatrix} \sqrt{2\over 3} &1\over \sqrt{3}&0 \\
				-1\over \sqrt{6}&1\over \sqrt{3}& 1\over \sqrt{2}\\
				-1\over \sqrt{6}&	1\over \sqrt{3}& -1\over \sqrt{2}
				
\end{pmatrix} .
\label{eq:HPS}
\end{equation}
\begin{wrapfigure}[14]{l}{.6\textwidth}
\centering
\vspace{-1cm}
\includegraphics*[scale=.32]{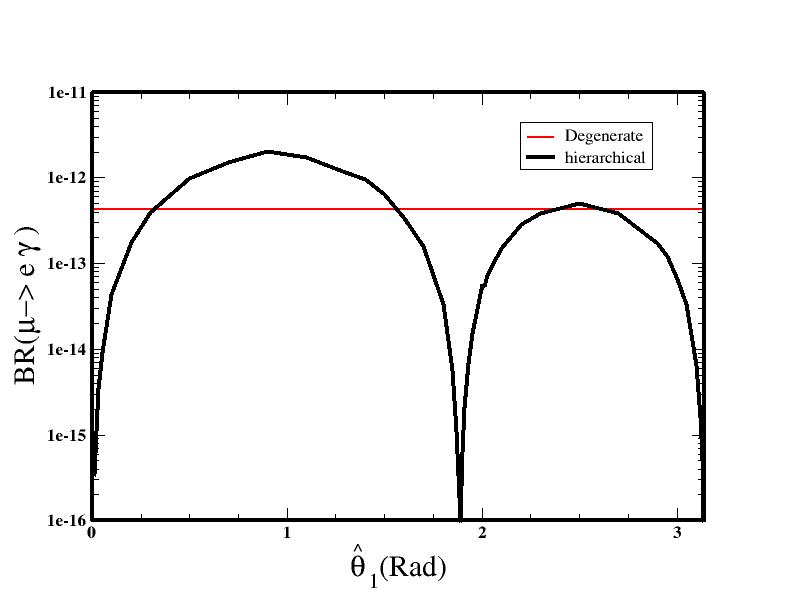}
\vspace{-.7 true cm}
\caption{\it $BR(\mu \rightarrow e \gamma)$ vs $\hat\theta_1$ under two different assumptions for the right-handed neutrinos using the CMSSM with $m_0=650$~GeV, $m_{1/2}=700$~GeV, $A_0=-1400$~GeV and $\tan\beta=40$, $\mu>0$ and $M_3=2.5\cdot 10^{12}$~GeV.}
\label{fig:mueg_hd}
\end{wrapfigure}  
\vspace{.7 true cm}

In order to determine the slepton mixing parameters, we need a
specific form of the product $Y_\nu^\dagger Y_\nu$,  shown in
(\ref{offdiagonal}). Even with the assumption of hierarchical light
neutrinos and a fixed $U$ matrix we can still have different
predictions for BR($l_j \rightarrow l_i + \gamma)$ depending on the
model used for RH neutrino masses. For instance, in the case of SU(5)
models several examples are provided in ref. \cite{Cannoni:2013gq}. The case
of  degenerate RH neutrinos implies hierarchical  $Y_\nu$ matrices,
since they inherit the neutrino mass hierarchy, while the BRs are
independent of the matrix $O_R$. In the case of hierarchical RH
neutrinos, the matrix $Y_\nu$ can have large entries even for the
first two generations, increasing the predicted BRs. However, the
matrix $O_R$ may induce large cancellations that can result in lower BRs with larger RH neutrino masses and couplings in the case of degenerate RH neutrino masses.  This behavior can be understood by comparing the predictions for   $BR(\mu \rightarrow e \gamma)$ in the case of degenerate and hierarchical RH neutrinos. 

 In the case of degenerate right-handed neutrinos ($M_i=M_N$): 
 \begin{equation}
 Y_\nu= Y_0 \sqrt{m_{\nu}^\delta \over m_{\nu_3}} U^\dagger,
 \end{equation}
 
 \noindent where $Y_0 =\sqrt{2}/{v_u}  \sqrt{M_N \cdot m_{\nu_3}}$. For the hierarchical case, assuming that $m_{\nu_1}\sim 0$ and that $M_1, M_2<<M_3$, using $U$ values from equation (\ref{eq:HPS}) and 
 the generic rotation (\ref{eq:or}), we find that $Y_\nu$ only depends on $\hat{\theta}_1$: 
 
 \begin{equation}
 Y_{\nu}= \bar{Y}_0 \begin{pmatrix} 0&0&0\\
 0&0&0\\
 \sqrt{m_{\nu_2} \over {m_{\nu_3}}}   \cdot \frac{\hat{s}_1} {\sqrt{3}}& \frac{\hat{c}_1} {\sqrt{2}}+ 
 \sqrt{m_{\nu_2} \over {m_{\nu_3}}}   \cdot \frac{\hat{s}_1} {\sqrt{3}}&
 - \frac{\hat{c}_1} {\sqrt{2}}+ 
 \sqrt{m_{\nu_2} \over {m_{\nu_3}}}   \cdot \frac{\hat{s}_1} {\sqrt{3}}
 \end{pmatrix} .
 \label{eq:hierarchical}
 \end{equation}
  where $Y_0 =\sqrt{2}/{v_u}  \sqrt{M_3 \cdot m_{\nu_3}}\cdot \hat{c}_2$.

Figure \ref{fig:mueg_hd}   shows the prediction for  $BR(\mu \rightarrow e \gamma)$ under
both assumptions for the right-handed neutrinos, assuming the CMSSM with $m_0=650$~GeV, $m_{1/2}=700$~GeV, $A_0=-1400$~GeV and $\tan\beta=40$, $\mu>0$ and $M_3=2.5\cdot 10^{12}$~GeV. For the case of hierarchical right-handed neutrinos we assume $\cos(\hat\theta_2)=1$. 
It is easy to conclude that assuming hierarchical light neutrinos and a common
scale for the right-handed neutrinos provides a simple benchmark. 
In this case, using (\ref{eq:casas}), we find 
\begin{equation}
Y_\nu^\dagger Y_\nu= \frac{2}{v_u^2}M_R U m_\nu^\delta U^\dagger~.
\label{eq:ynu2}
\end{equation}
Under the assumption of common masses for the heavy Majorana neutrinos, the LFV effects are independent of the matrix $O_R$. On the other hand, the predicted 
curve of the hierachical case in Figure \ref{fig:mueg_hd} depends on $\hat\theta_1$  as $\left({Y_{\nu}}^{\dagger} Y_{\nu}\right)_{12}^2$,  
where $ Y_{\nu}$ is given in eq.~(\ref{eq:hierarchical}).

\begin{table}[t!]
\begin{center}
\begin{tabular}{|c|ccc|}
\hline  
\rule[-2ex]{0pt}{5.ex} \bf{SUSY parameters} & 4-2-2  & SU(5)  & FSU(5)
\\ 
\hline
\rule[-2ex]{0pt}{3.ex} $100\;\text{GeV}\leq  m_0 \leq 10\; \text{TeV}$  & $0\leq x_u \leq 2$ & $0\leq x_u \leq 2$ & $0\leq x_u \leq 2$ 
\\ 
\rule[-2ex]{0pt}{3.ex} $50 \; \text{GeV}\leq  m_{1/2} ~(M_2$ ~in~4-2-2) $\leq 10$\; \text{TeV} & $0\leq x_d \leq 2$ & $0\leq x_d \leq 2$ & $0\leq x_d \leq 2$ 
\\
\rule[-2ex]{0pt}{3.ex} $-10\;  \text{TeV}\leq  A_{0} \leq 10\; \text{TeV}$  &  $0\leq x_{LR} \leq 2$  & $0\leq x_5 \leq 2$ & $0\leq x_5 \leq 2$ 
\\ 
\rule[-2ex]{0pt}{3.ex} $2\leq \tan\beta \leq 65$ &  &  &  $0\leq x_R \leq 2$ 
\\
\rule[-2ex]{0pt}{3.ex} $-3000\; \text{GeV}\leq  M_3$ (in~4-2-2) $\leq 10\; \text{TeV}$ & & & 
\\   
\hline 
\end{tabular}
\caption{\it Parameter ranges sampled in our scan of the parameter spaces of the GUT models we study.}
\label{tab1}
\end{center}
\end{table}

\section{LFV, Dark Matter and the LHC}

We perform parameter space scans similar to those in~\cite{EGLR, Gomez:2018zzw,Gomez:2018efz}, where the initial 
conditions of the soft terms are determined by a unification group that breaks 
at $M_{GUT}$ (defined as the scale where the $g_1$ and $g_2$ couplings meet, 
while $g_3(M_{GUT})$ is obtained by requiring $\alpha_s(M_Z)=0.1187$).  
For our analysis we use the {\tt Superbayes}
\cite{Bertone:2011nj,Strege:2012bt,Bertone:2015tza}, 
package to perform 
statistical inference of SUSY models, which is linked to {\tt SoftSusy} \cite{softsusy} 
to compute the SUSY spectrum, to {\tt MicrOMEGAs} \cite{MicrOMEGAs} and {\tt DarkSUSY} \cite{DarkSUSY} 
to compute DM observables, to {\tt SuperIso} \cite{SuperIso}
to compute flavour physics and the muon anomalous magnetic moment $g-2$.
The {\tt MultiNest} \cite{multinest} algorithm is used to scan the
parameter space and identify regions compatible with the data.

We have scanned the parameter spaces of the three GUT groups over the
broad ranges of parameters shown in Table~\ref{tab1}, including soft SUSY-breaking terms up to 10 TeV,
with the results that we now discuss. 
In addition to the dark matter density constraint
mentioned above (\ref{DMdensity}), we impose the
following constraints:
\begin{equation}
 123~{\rm GeV}\leq m_h\leq 127~{\rm GeV} \, ,
\label{mh}
\end{equation}
which includes an allowance for the theoretical
uncertainty in the calculation of $m_h$ in the CMSSM, which is computed using \cite{softsusy}.
We extract the following B-physics constraints from \cite{Amhis:2019ckw}:
\begin{equation}
 1.1\times 10^{-9} \leq{\rm BR}(B_s \rightarrow \mu^+ \mu^-)
  \leq 6.2 \times10^{-9} \, ,
  \label{bsmumu}
  \end{equation}
which accommodates the range allowed 
experimentally at the 2-$\sigma$ level,
\begin{equation}
2.99 \times 10^{-4} \leq
  {\rm BR}(B \rightarrow X_{s} \gamma)
  \leq 3.87 \times 10^{-4} \, ,
\label{bsgamma}
\end{equation}
which also covers the 2-$\sigma$ experimental 
range, and
\begin{equation}
0.15 \leq \dfrac{
 {\rm BR}(B_u\rightarrow\tau \nu_{\tau})_{\rm MSSM}}
 {{\rm BR}(B_u\rightarrow \tau \nu_{\tau})_{\rm SM}}
        \leq 2.41 \, ,
\label{Btaunu}
\end{equation}
which covers the 3-$\sigma$ experimental range.
These constraints are implemented as described in \cite{Bertone:2015tza}.

In addition, we impose the constraints on the
spin-independent (SI) neutralino-nucleon cross-section
provided by the LUX~\cite{LUX}, Xenon-1T~\cite{XENON1T_new} and PandaX experiments~\cite{Cui:2017nnn}.

\begin{figure}
\includegraphics*[scale=.2]{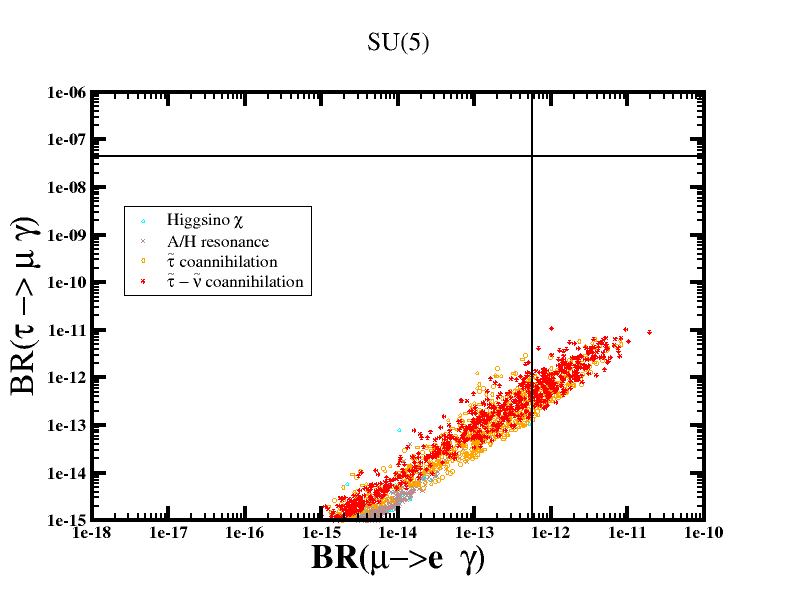}
\includegraphics*[scale=.2]{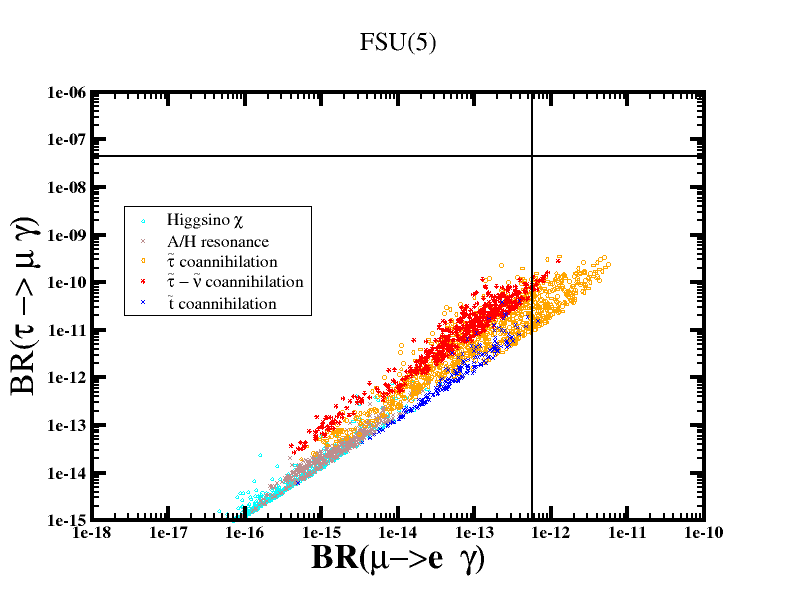}
\includegraphics*[scale=.2]{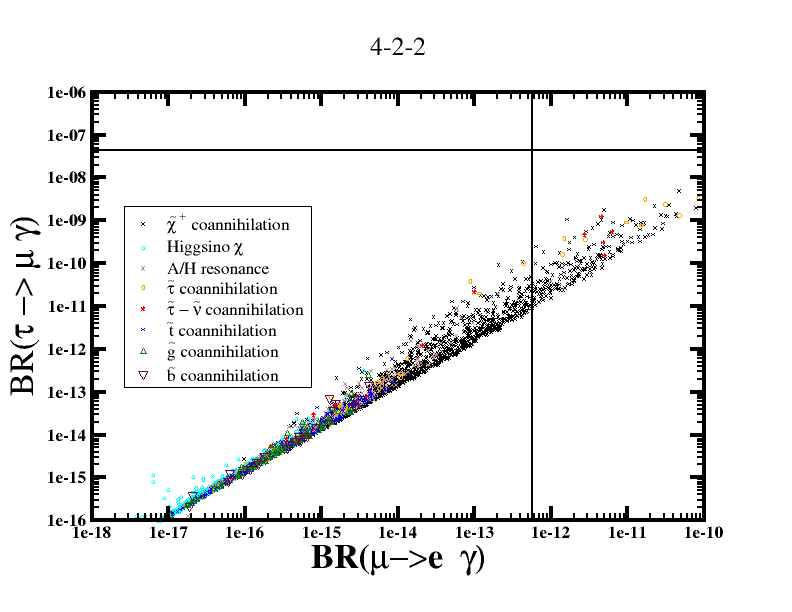}
\includegraphics*[scale=.2]{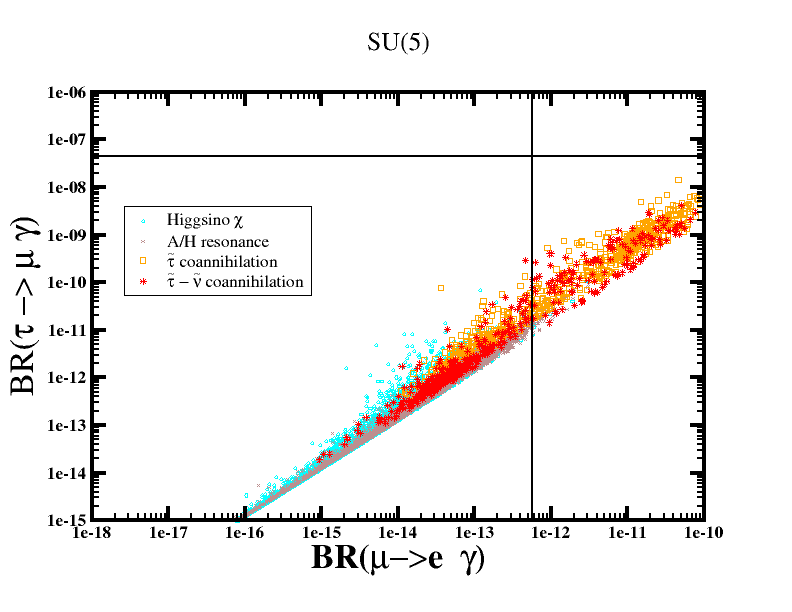}
\includegraphics*[scale=.2]{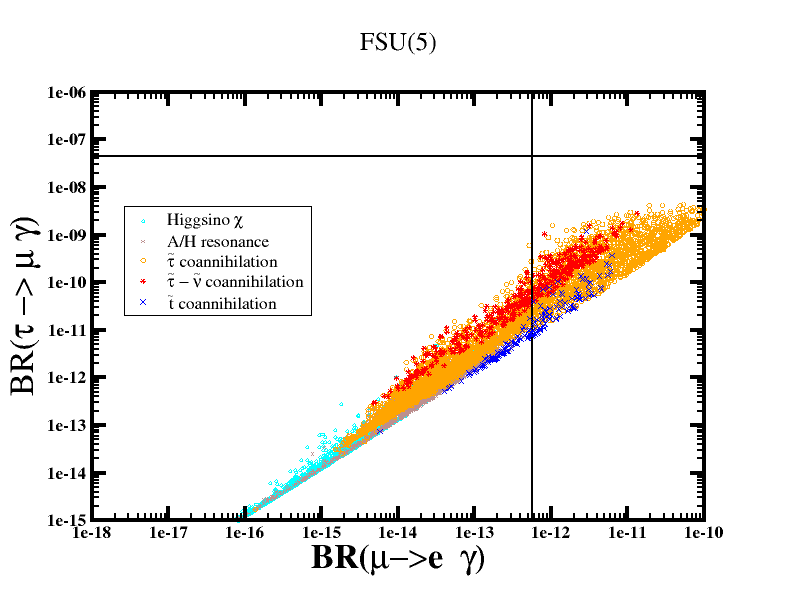}
\includegraphics*[scale=.2]{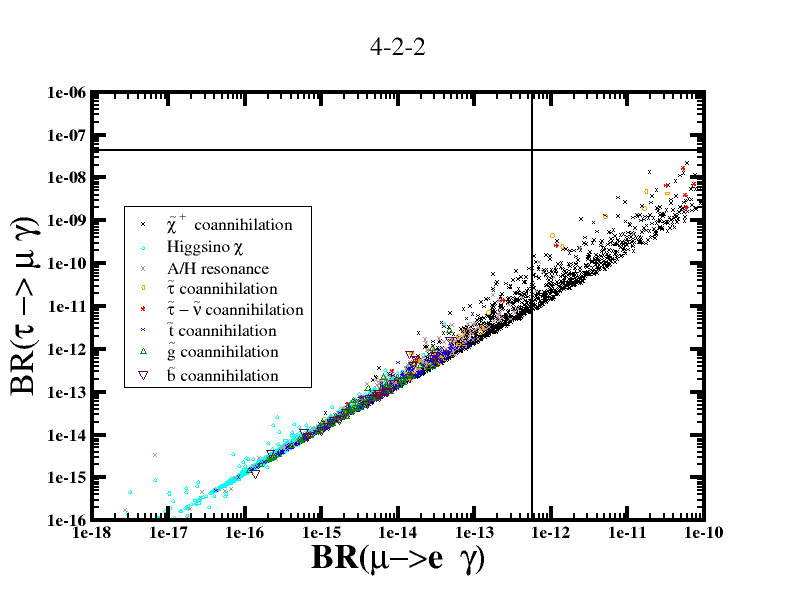}
\vspace{-1.0 true cm}
\caption{\it Predictions for ${\rm BR}(\tau \rightarrow \mu \gamma)$  and
  ${\rm BR}(\mu \rightarrow e \gamma)$ for (from left to right) the SU(5), FSU(5) and 4-2-2 models.
  The upper (lower)  panels  assume $M_N=2.5\cdot 10^{12}$ ($M_N=10^{13}$~GeV). 
  The symbols correspond to classes of models representing the
  DM scenarios described in the text, which are indicated in the plot legends.}
\vspace{0.5 true cm}
\label{fig:teg-megIII}
\end{figure}

%


\subsection{\boldmath{${\rm BR}(l_i \rightarrow l_j \gamma)$}}

The processes $l_i\rightarrow l_j \gamma$ with $i\neq j$ are allowed
at potentially observable levels in SUSY models with flavour mixing among leptons and
their scalar partners. In the CMSSM this mixing does not occur, due to
the assumption of universal soft terms at the GUT
scale. However, this simple SUSY extension of the SM cannot explain
neutrino flavour oscillations and, when the model is supplemented with a
mechanism to account for them, flavour oscillations of charged leptons also
occur.  In the MSSM supplemented by a
a type-I see-saw as described in the previous Section, which is compatible with
the available neutrino data, the uncertainties in the latter may lead to
LFV predictions that can differ by several orders of magnitude. 
Our target in this work, therefore, is not only to analyze the possibility of
observing  LFV in current experiments, but also to understand the impact of the 
bounds on BR($\mu \to e \gamma)$ on the perspectives for LHC data.
In the simplified see-saw scenario presented in Section 4, we must still
specify the following parameters: 

\begin{itemize}
\item The right-handed neutrino scale, which we assume to be common for all generations. 
\item Lepton-slepton mixings parametrized by a matrix similar to the
  PMNS matrix at the GUT scale. We fix the entries assuming that they are real
  and that their values are such that the neutrino observables are
  predicted at their experimental central values.  
 In this simple scheme, the product  $Y_\nu^\dagger Y_\nu$ is defined by the PNMS matrix as in (\ref{eq:ynu2}).
\item SUSY soft masses are flavour-independent at the GUT scale. However, we allow the sfermions belonging to different representations of the unification group to have different soft masses. 
\end{itemize}


The first two points are discussed in this Section, as illustrated in
Fig.~\ref{fig:teg-megIII}, whereas
the third point requires a more elaborate treatment,
and we dedicate the two subsequent Sections to it. 

For fixed light neutrino masses, Eq.~(\ref{meff}) links the ratio of the
square of the Yukawa couplings to the right-handed neutrino masses. We see that higher right-handed
neutrino masses imply, in general, higher Yukawa couplings and larger
mixings of the scalar sleptons in Eq.~(\ref{offdiagonal}), and hence larger
LFV branching ratios (BRs).  Fig.~\ref{fig:teg-megIII} compares model predictions with the 
experimental upper limits on BR($\tau \to \mu \gamma)$ and BR($\mu \to e
\gamma)$. Comparing the upper and lower panels, we can understand
how an increase in $M_R$ by a factor of 4 would imply the exclusion of many
models by the current bound on BR($\mu \to e \gamma)$. For the rest
of  our analysis, we use $M_R = 2.5 \cdot 10^{12}$~GeV.
In this case, most of the points that can be explored at the LHC will predict BR($\mu \to e \gamma)$ 
between the current upper limit and a possible future sensitivity one order of magnitude lower.  
\begin{figure}
\begin{center}
\includegraphics*[scale=.2]{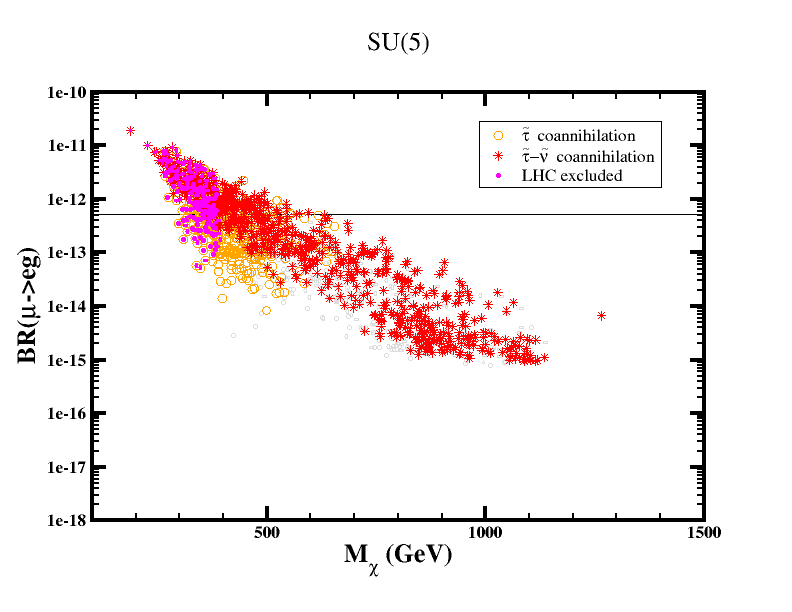}
\includegraphics*[scale=.2]{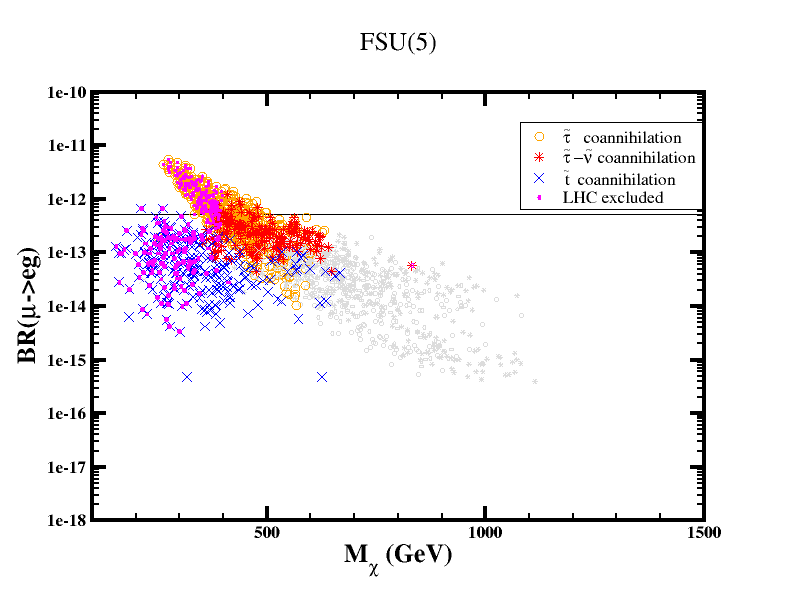}
\includegraphics*[scale=.2]{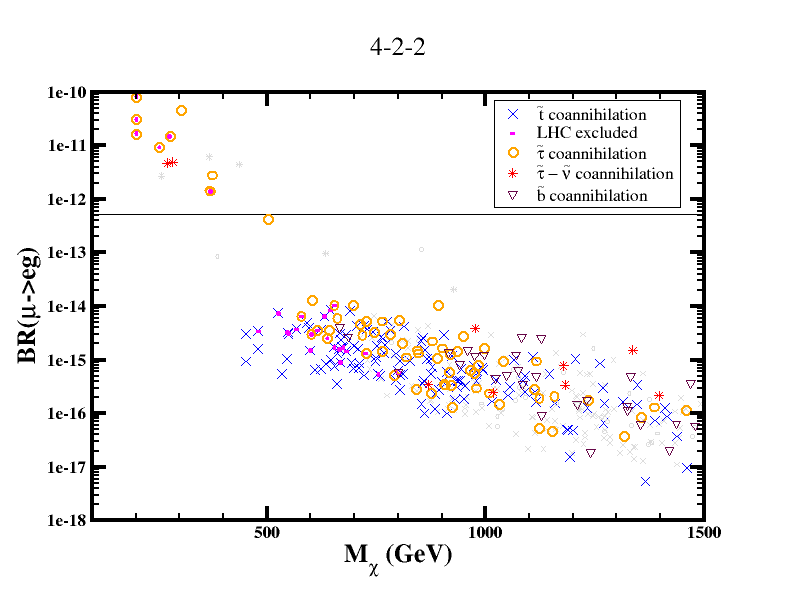}\\
\vspace{-.50 true cm}
\caption{\it Prediction for  ${\rm BR}(\mu \rightarrow e \gamma)$  vs
  $m_\chi$ for (from left to right) the SU(5), the FSU(5) and 4-2-2 models, 
  in scenarios where sfermions coannihilate with the LSP.
 We use the same notation for the DM models
 as in Fig.~\ref{fig:teg-megIII}. Models with parameters not
  detectable at the 
LHC are marked in grey, while excluded models are marked with purple
dots. We assume $M_N=2.5\cdot 10^{12}$~GeV in all three cases.}
\label{fig:br_457}
\end{center}
\end{figure}

The correlation  between BR($\tau \to \mu \gamma)$ and  BR($\mu \to e
\gamma)$ is almost linear, with the prediction of the first being
larger than the second by a factor of 10, while the experimental
bounds are five orders of magnitude apart. In our study we fixed $Y_\nu^\dagger Y_\nu$
from the PMNS matrix, requiring common right-handed neutrino masses. Although
this cannot be considered general, the values of LFV tau decays are
maximized by large 2-3 mixing in the PNMS matrix. Furthermore, in  SU(5)
group symmetries can relate the PMNS and Cabibbo-Kobayashi-Maskawa (CKM) matrix,
leading to large mixing in the 2-3 sector. It is nevertheless possible to
find particular textures for $Y_\nu$ and $M_N$ for which the
ratios of $\mu$ and $\tau$ decays are simultaneously closer to the
experimental bounds. These cases will, however, typically imply smaller
values for the Dirac Yukawa couplings of the first and second generations,
predicting less restrictive BRs. Our study can be considered as
targeting the kinds of textures that predict large charged LFV. 

\subsection{Combining $\mu \rightarrow e \gamma$ and LHC bounds}
\begin{figure}
\begin{center}
\includegraphics*[scale=.2]{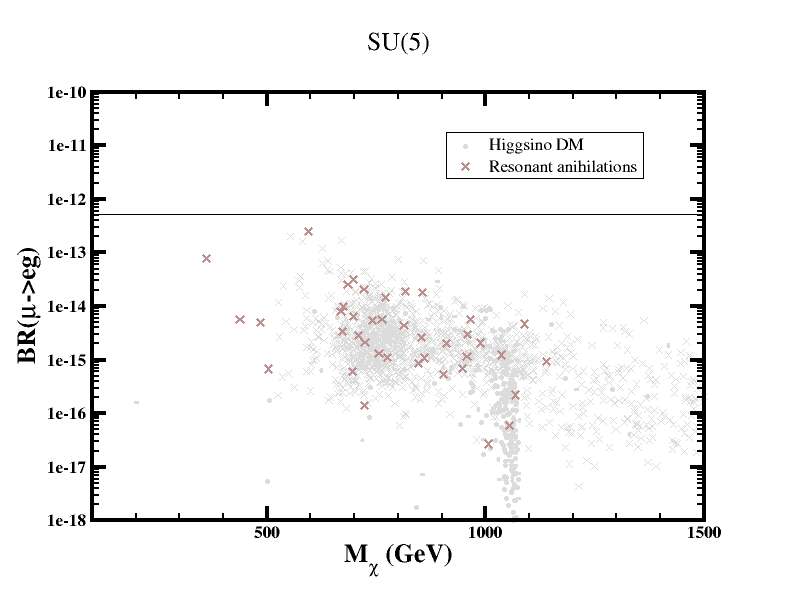}
\includegraphics*[scale=.2]{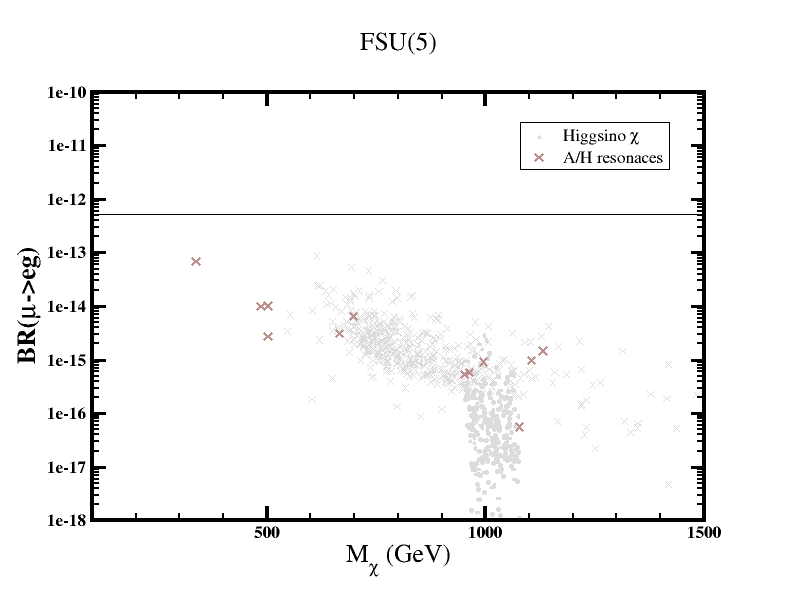}
\includegraphics*[scale=.2]{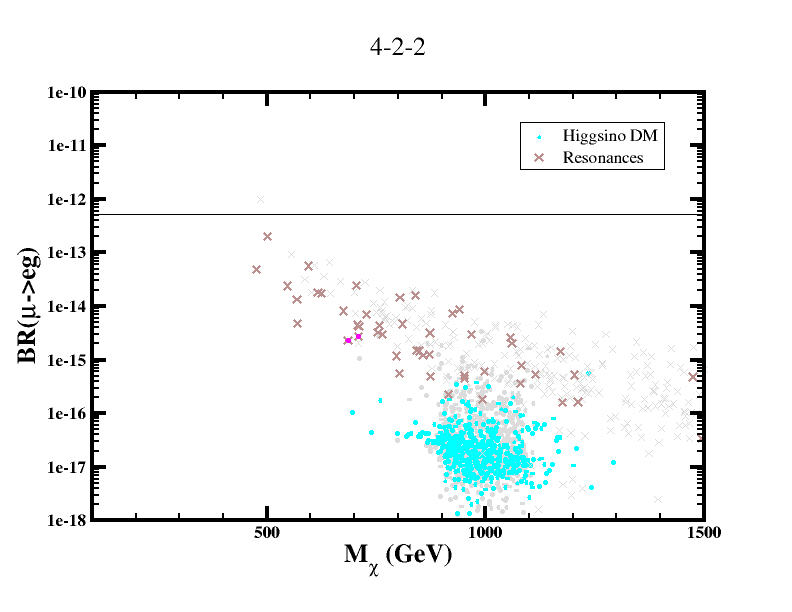}\\
\vspace{-0.5 true cm}
\caption{\it As in Fig.~\ref{fig:br_457}, for Higgsino DM and for models of resonant coannihilation.}
\label{fig:br_13}
\end{center}
\end{figure}

The scale  $M_R = 2.5 \cdot 10^{12}$ GeV was chosen as representative. It also
turns out that no points are excluded by $\tau\rightarrow \mu
\gamma$, since $\mu \rightarrow e \gamma$ is more
restrictive. This bound, in combination with large mixing for solar and
atmospheric neutrinos, 
also excludes models with rare $\tau$ decays
at the levels of the experimental limits in almost all
natural textures. Keeping this in mind, we proceed to analyze the
predictions for this decay in different unification schemes, studying all
kinds of DM models.  Since the signal largely depends on the SUSY
particle spectroscopy, we combine our analysis with consideration of the LHC data for the specific unified SUSY models under consideration.  For this purpose we follow a similar procedure as that applied in Refs.~\cite{Gomez:2018zzw,Gomez:2018efz}.  Each model can be associated to a particular set of particle mass hierarchies and decays, which are then compared with the generic analyses provided by the ATLAS and CMS collaborations \cite{Okawa:2011xg,Chatrchyan:2013sza}. These comparisons are made with the help of Simplified Model Spectra (SMS) which can be defined by a set of hypothetical SUSY particle masses and a sequence of decay patterns that have to be compared with those expected in any specific model. An individual check has to be done for every model, while, due to mismatches between the theoretical predictions and the experimental analyses, it is not possible to provide contour plots where one can easily see which mass ranges are excluded. This task is simplified by using public packages like {\tt Smodels-v1.2.2}~\cite{Ambrogi:2018ujg}, which provides a powerful tool for performing a fast analysis of a large number of models~\cite{Kraml:2013mwa,Ambrogi:2017lov}. Using this package, the theoretical models are mapped onto SMS and can be compared with the existing LHC bounds if there is a match in the respective topologies. In each model the mass spectrum is generated using {\tt SoftSusy} and the corresponding decay ratios are calculated using {\tt SUSY-HIT}~\cite{Djouadi:2006bz}. The cross-section information is then inserted in {\tt Smodels-v1.2.2} through a call to {\tt Pythia~8.2}~\cite{Sjostrand:2014zea}.

We classify the models as follows, according to their LHC prospects:\\
(i) ~~Those that  are excluded by the current LHC bounds; \\
(ii) ~Those that can be compared with the LHC data  and are not excluded; \\
(iii) Those that cannot be tested at the LHC, i.e., points that predict either processes with very low cross sections or topologies that are not tested at the LHC. \\
~~\\
In Fig.~\ref{fig:br_457} and following figures, we denote points of the same DM class with the same symbol as in Fig. \ref{fig:teg-megIII} but changing the colour according to the LHC prospects of the model:  
Points of categories (i) and (ii) retain the same colour as in  Fig.~\ref{fig:teg-megIII}, adding a magenta dot for the excluded ones, whereas points of category (iii) are drawn in grey. 

We see in  Fig.~\ref{fig:br_457} that the  $\mu
\rightarrow e \gamma$ bound may be violated in DM models with coannihilations, whereas models with resonant annihilations and higgsino DM are not affected by this bound, as seen in Fig.~\ref{fig:br_13}. This can be attributed to the lighter masses of the sleptons in the coannihilation scenarios.  \\
~~\\
The LHC and charged LFV predictions of models populating classes of points
with different DM mechanisms can be compared in the different unification scenarios: \\
~~\\
\begin{wrapfigure}[14]{l}{.6\textwidth}
\centering
\vspace{-1cm}
\includegraphics*[scale=.32]{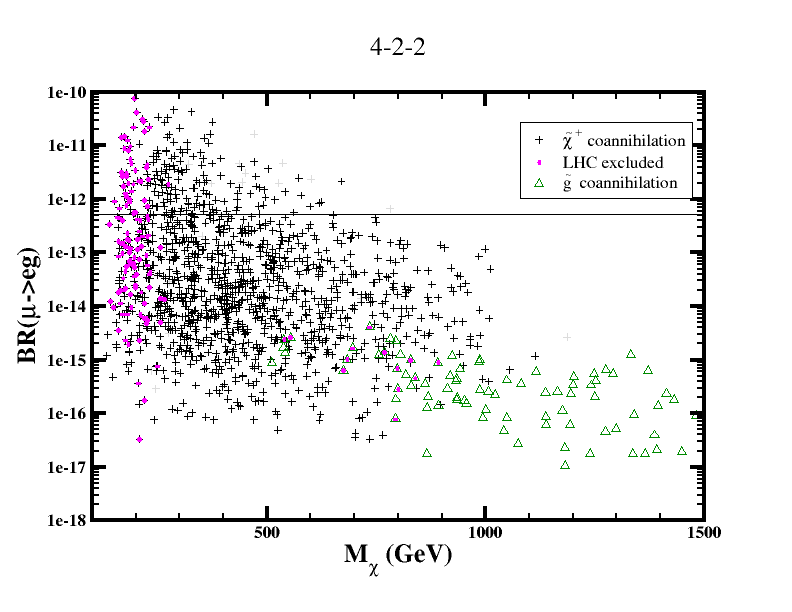}
\vspace{-.5 true cm}
\caption{\it As in Figs.~\ref{fig:br_457}  and \ref{fig:br_13} for models with chargino and gluino coannihilations in the 4-2-2 GUT.}
\label{fig:br_26}
\end{wrapfigure}
{\bf $\tilde{\tau}-\chi$ and  $\tilde{\tau}-\tilde{\nu}-\chi$
  coannihilation:}  These mechanisms are particularly interesting, since
  they both predict LFV and LHC signals within experimental
  reach. In the  $\tilde{\tau}-\chi$ scenario, the lighter stau
  is determined by left-right mixing, and the  $\tilde{\tau}-\tilde{\nu}-\chi$ is
  the limiting case where the $\tilde{\tau_1}$ is mainly left-handed. 
  We should also take into account the fact that LFV is induced
  mainly in the left-left sector of the slepton mass matrix,  due to the see-saw mechanism, therefore models with larger left-stau composition and smaller masses tend to have larger LFV decay rates. As seen in Fig.~\ref{fig:br_13}, this scenario is very interesting in SU(5) and FSU(5), since these models predict both LFV and LHC signals within experimental reach. Models with LSP 
masses above ~400 GeV are not excluded in either scenario,  
but the different representation assignments and hence soft masses change the slepton
compositions, with manifest implications for the LFV predictions, which are specific for
each group. For instance, whereas in SU(5) most of the points with 
$\tilde{\tau}-\tilde{\nu}-\chi$ coannihilations violate the experimental bound, in
FSU(5) they are still allowed. In the case of 4-2-2 models, there is a
left-right splitting of the sfermion soft masses, implying that 
points with stau coannihilations are more difficult to find than in
SU(5), as can be seen in the corresponding panel  of
Fig.\ref{fig:br_457}. Moreover, due to gaugino mass relations, the
charginos and neutralinos can be heavier than in SU(5) models,
leading to lower  BR( $\mu \rightarrow e \gamma$). \\
~~\\
{\bf $\tilde{t}-\chi$ coannihilation:} Fig.~\ref{fig:br_13} shows that such models are present in
   the FSU(5) and in the 4-2-2 schemes, but the predictions are different in the two frameworks. In FSU(5), models with LSP masses up to 700 GeV can predict ratios up to one order of magnitude below the current bound, whereas in 4-2-2 models the LSP mass can be  larger, with BR($\mu \rightarrow e \gamma$) two orders of magnitude below the experimental limit.\\
~~\\
{\bf $A/H$ resonances:}  As can be seen in
  Fig.~\ref{fig:br_13}, the predictions for LFV decays are below the
  current limits. However, there are some differences between the three
  GUTs for the points with good prospects for both the LHC and
  BR($\mu \rightarrow e \gamma$), which are easier to 
find in 4-2-2 and SU(5) than in FSU(5). \\
~~\\
{\bf Higgsino DM:} Fig.~\ref{fig:br_13} shows that this class
  of points does not predict charged LFV of experimental interest, due to the
  heavy SUSY masses in the three GUT schemes; these models are also
  out the LHC reach in all SU(5) cases. However, the LSP composition
  is different in the three schemes;  for instance, in the 4-2-2 model the LSP
  is almost a pure Higgsino and, even if BR( $\mu \rightarrow e \gamma$) is low,
  some model points can be tested at the LHC. \\
~~\\
{\bf $\tilde{\chi}^+-\chi$ and  $\tilde{g}-\chi$ coannihilations:} These DM
  classes appear only in the 4-2-2 case, due to its GUT relation on gaugino masses. 
  As can be seen in Fig.~\ref{fig:br_26}, models with $\tilde{\chi}^+-\chi$  coannihilations have good detection prospects for both LFV decays and at the LHC. Points with   $\tilde{g}-\chi$ 
  coannihilation are still within the LHC reach, while the BR($\mu \rightarrow e \gamma$) 
  predictions are low.\\

\begin{figure}
\begin{center}
\includegraphics*[scale=.3]{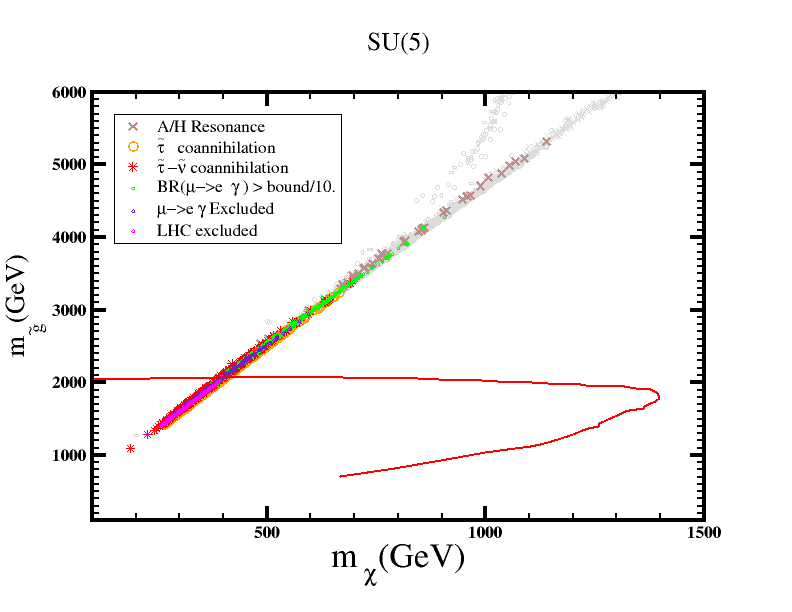}
\includegraphics*[scale=.3]{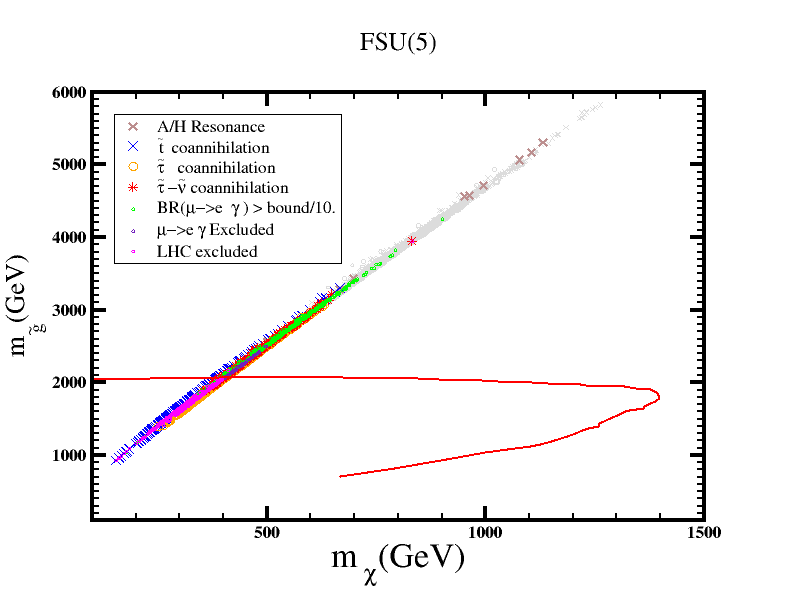}\\
\vspace{-0.5 true cm}
\includegraphics*[scale=.3]{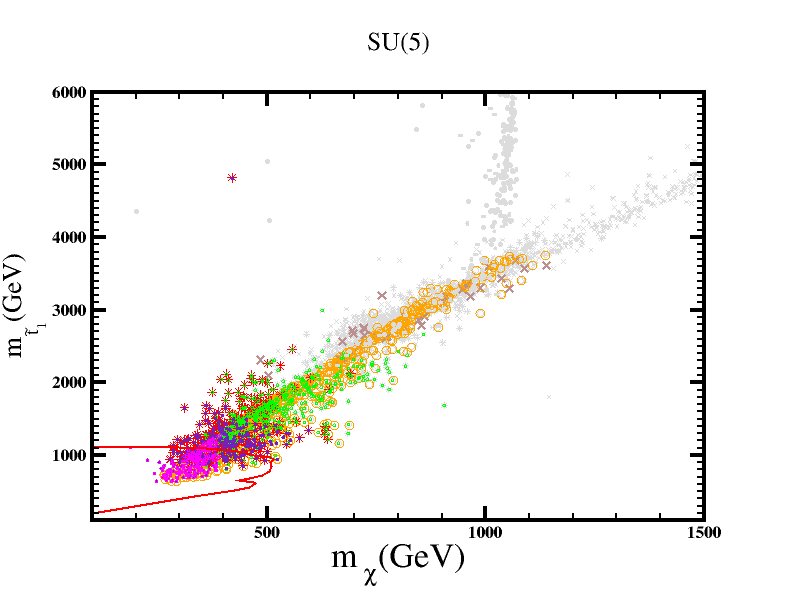}
\includegraphics*[scale=.3]{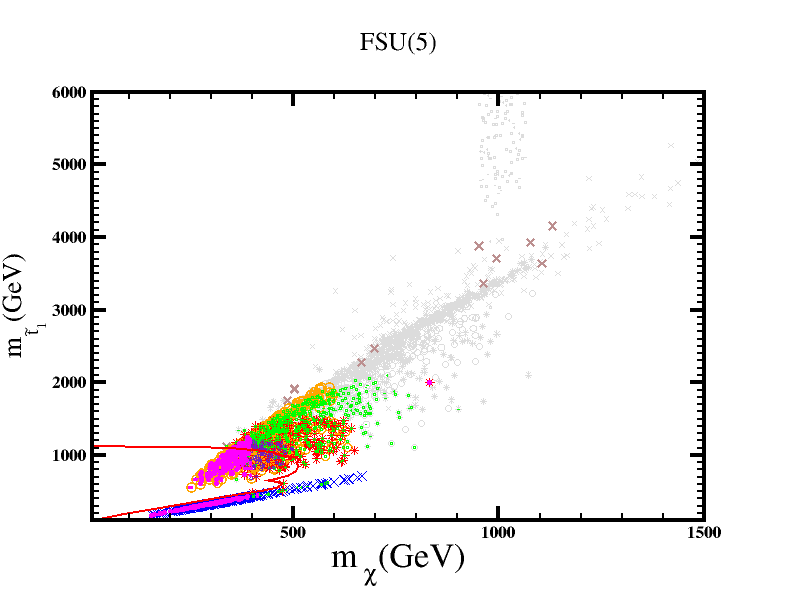}\\
\vspace{-0.5 true cm}
\includegraphics*[scale=.3]{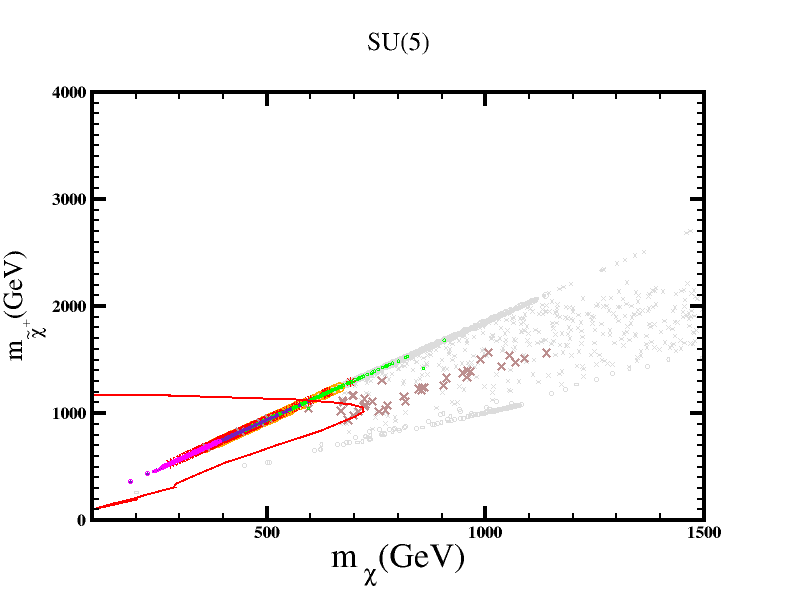}
\includegraphics*[scale=.3]{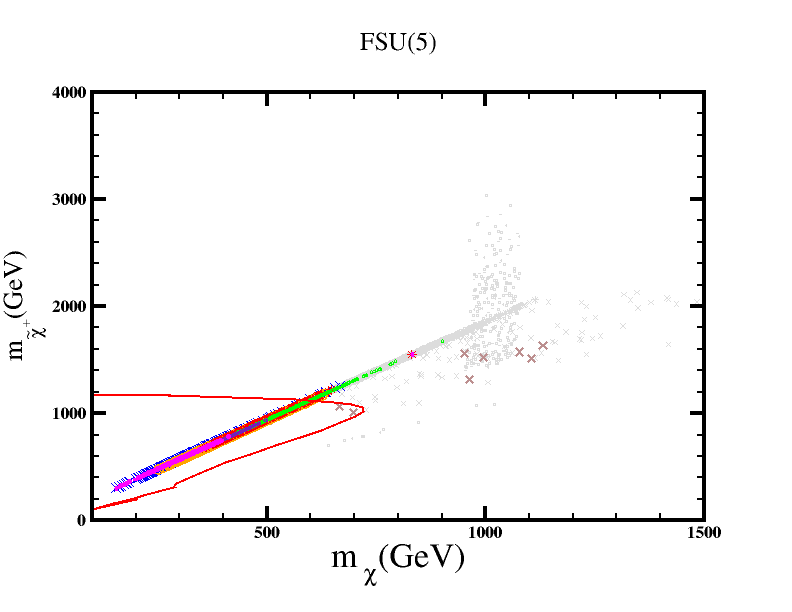}\\

\caption{\it LHC prospects for the SU5 and FSU5 models. The points follow the
  notation of Figs.~\ref{fig:br_457}, \ref{fig:br_13} and
  \ref{fig:br_26}. The meanings of the solid red lines are explained in the
  text. Indigo crosses indicate points excluded by the limit on BR($\mu
  \rightarrow e \gamma$),  whereas the green crosses mark points that lie
between the current bound and one order of magnitude below it.}
\label{fig:mgmchi_fsu5}
\end{center}
\end{figure}


\begin{figure}
\begin{center}
\includegraphics*[scale=.3]{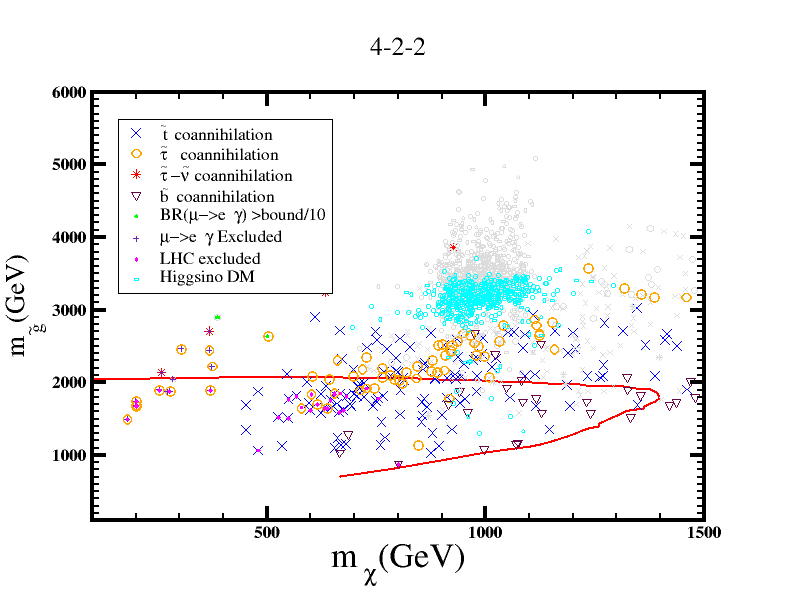}
\includegraphics*[scale=.3]{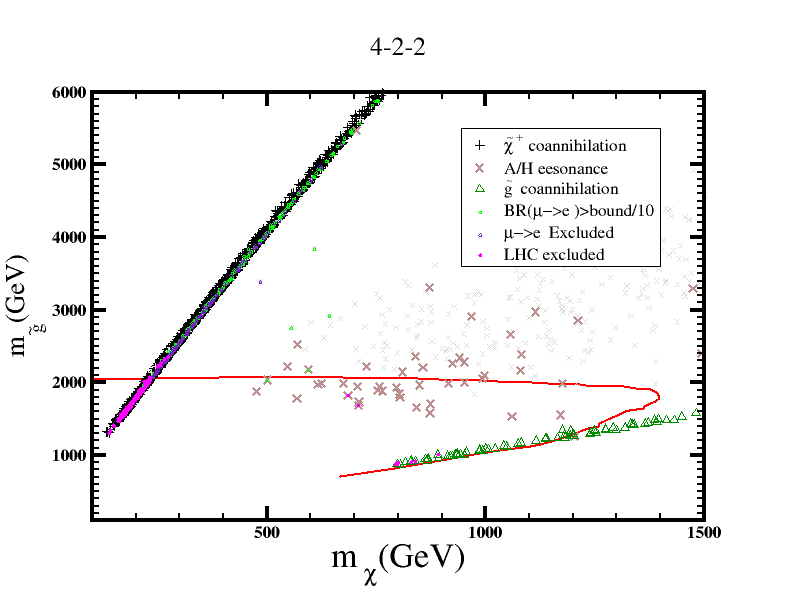}\\
\vspace{-0.5 true cm}
\includegraphics*[scale=.3]{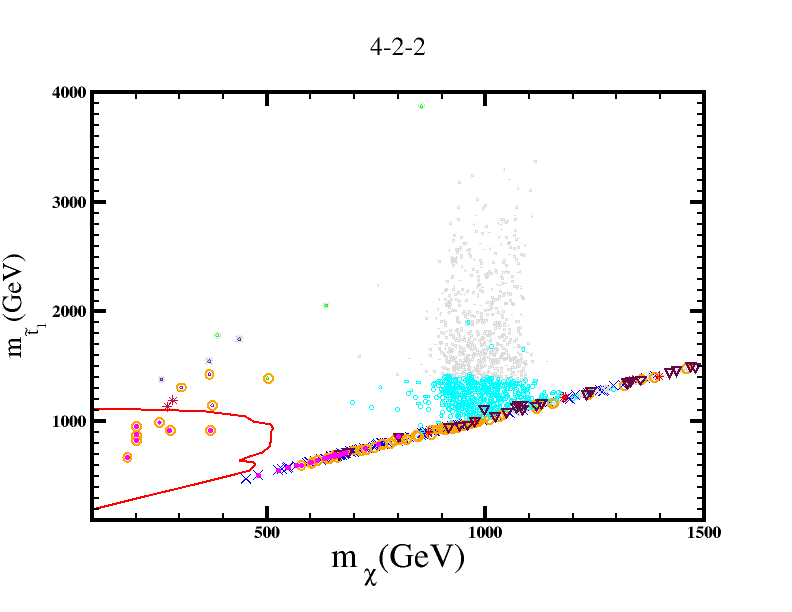}
\includegraphics*[scale=.3]{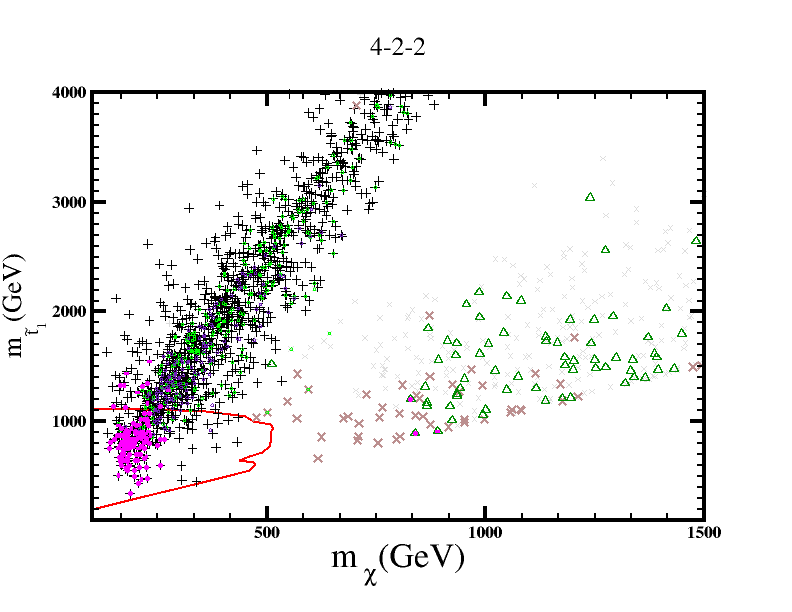}\\
\vspace{-0.5 true cm}
\includegraphics*[scale=.3]{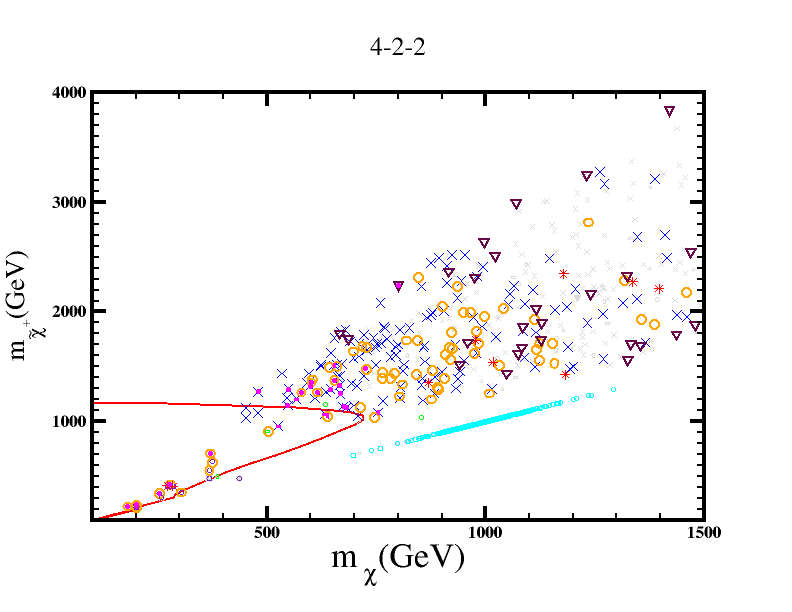}
\includegraphics*[scale=.3]{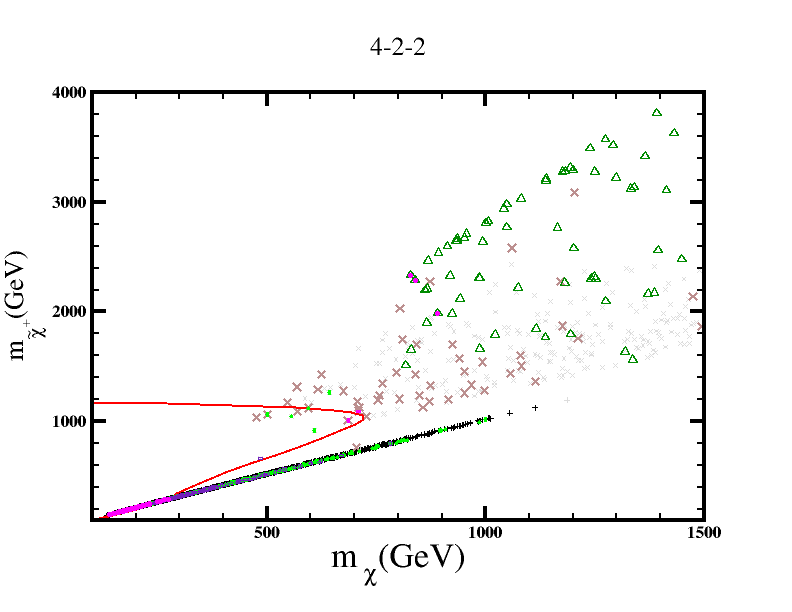}\\

\caption{\it LHC prospects for the 4-2-2 model, following the notations of
  Fig. \ref{fig:mgmchi_fsu5}. For clarity of presentation, in the left
  panels we display predictions for models with sfermion
  coannihilations,  whereas in the right panels we display the remaining cases.}
\label{fig:mgmchi_ps}
\end{center}
\end{figure}

\subsection{LFV signals, SUSY spectroscopy and DM detection}

In this Section we discuss the LHC prospects for discovering SUSY
combined with a possible charged LFV signal.  
The results are shown in Figs.~\ref{fig:mgmchi_fsu5} and \ref{fig:mgmchi_ps}, which 
plot  SUSY particle  masses vs. $m_\chi$, in order 
to compare directly the range of SUSY masses to which
the LHC is sensitive with those that give rise to detectable LFV signatures.
In the case of SU(5) and FSU(5), each panel  contains all classes of points, while in the 4-2-2 case the
different classes are shown in two panels, for
clarity of presentation. 

We follow the same notation as in the previous Section, with purple dots denoting points excluded by the LHC. In addition to the symbols introduced in the previous Sections, we introduce two more, to show the impact of the LFV predictions on the SUSY spectrum: \\
- Indigo crosses mark points excluded
by the current bound on BR($\mu \rightarrow e \gamma$), and \\
- Green crosses mark points with predictions for BR($\mu \rightarrow e \gamma$)
between the present bound and a factor of 10 below this value.\\
In addition, the solid red lines are obtained by combining the simplified model bounds from LHC searches. 
Since these bounds often do not apply directly to our particular cases, 
this boundary should not be considered as an exclusion line, though excluded points would lie within at least one of these contours. Nevertheless, it is useful to include this line for illustrative
purposes, since it gives 
an idea  of the range of masses explored at the LHC for every SUSY
particle.  
  
The upper panels in Fig~\ref{fig:mgmchi_fsu5} and \ref{fig:mgmchi_ps} display LHC and LFV results on 
 $m_{\tilde{g}}-m_\chi$ contour plots. Since in SU(5) and FSU(5) we assume universal gaugino masses at the GUT scale, 
 all except the Higgsino DM models lie on the proportionality
 lines obtained from the GUT relations. Among other relations, the
 neutralino mass is in general proportional to that of the gluino, something that does not hold in 4-2-2 where,
 due to its different group structure, the distribution of models (shown in Fig.~\ref{fig:mgmchi_ps}), follows different patterns.
 The sfermion coannihilation cases (left panel) do not show any correlation in the $m_{\tilde{g}}-m_\chi$ plane. The same holds for models with 
 Higgsino DM and with A/H resonances (right panel), 
 which deviate from the proportionality line.
 Chargino and gluino coannihiliations, on the other hand, display the pattern of mass correlations described in Ref.~\cite{Gomez:2018efz}.  We have checked that the excluded points inside the red contour in $m_{\tilde{g}}-m_\chi$ plots in   Figs.~\ref{fig:mgmchi_fsu5} and \ref{fig:mgmchi_ps} violate the constraint from the 0-lepton + jets +
$\cancel{\it{E}}_{T}$ channel \cite{CMS-SUS-16-033, CMS-SUS-16-036}. This bound affects all the models excluded by the LHC in SU(5), and most of the models excluded in the other two scenarios.

Although the superposition of models on  Figs.~\ref{fig:mgmchi_fsu5} and \ref{fig:mgmchi_ps} does not by itself allow a clear distinction among different DM scenarios, we can associate the excluded points to specific models by confronting these figures with the LFV predictions of Figs.~\ref{fig:br_457}, \ref{fig:br_13} and \ref{fig:br_26}. We see that models with sfermion coannihilations in SU(5) and FSU(5) are more affected by the LHC bounds than in the 4-2-2 case, especially for $\tilde{t}-\chi$ coannihilations. In all scenarios, LFV enables exploring a range of $m_{\tilde{g}}$ 
  far beyond the LHC bounds (up to about 4 TeV in SU(5) and FSU(5), and even larger values in 4-2-2 in the chargino coannihilation scenario). 
  
  The analysis of
  excluded models shown in the $m_{\tilde{t}}-m_\chi$ plots (middle panels of Figs.~\ref{fig:mgmchi_fsu5} and \ref{fig:mgmchi_ps}) indicates that they are affected by the bound due to searches for stop decays into $t-\chi^\pm$ \cite{ATLAS-SUSY-2013-02,CMS-SUS-16-049,CMS-SUS-16-050,CMS-SUS-16-051}. We see that the exclusion bound in the SU(5) and FSU(5) panels contains many points with slepton coannihilations, while points with $\tilde{t}-\chi$ coannihilations escape this bound. In the case of the 4-2-2 models shown in the left middle panel of Fig.~\ref{fig:mgmchi_ps}, we see that this bound is less effective for the same kind of models  than in the other GUTs. In the middle right panel of the same Figure, we see that the bound excludes many models with $\tilde{\chi}^\pm -\chi$ coannihilations with $m_\chi$ below 300 GeV. Regarding LFV, we see that the models present good detection prospects up to stop masses above 3 TeV (and even further in 4-2-2 models). However, in the case of 4-2-2, only models with $\tilde{\chi}^+-\chi$ coannihilation predict LFV within one order of magnitude of the current bound.

The  $m_{\tilde{\chi}^\pm}-m_\chi$ plane (bottom panels of Figs.~\ref{fig:mgmchi_fsu5} and \ref{fig:mgmchi_ps}) shows that it is possible to see models  excluded  by electroweak searches through the ATLAS multi-leptons + $\cancel{\it{E}}_{T}$ channel~\cite{ATL-CMS-multlep}. This channel is particularly important in models with $\tilde{\chi}^\pm-\chi$ coannihilation in the 4-2-2 scenario, where it can exclude models allowed by  0-lepton +jets + $\cancel{\it{E}}_{T}$. We see that most of the models with chargino mass $m _{\chi^\pm} \lesssim 300$ GeV are excluded by these searches. Regarding LFV in SU(5) and FSU(5), we see that the models give rise to good detection prospects for chargino masses up to 1.5 TeV whilst, in the case of 4-2-2, only models with chargino coannihilation present better prospects for LFV detection. In these cases, the masses reach the maximum value of 1 TeV within our data range. 
The impact is weaker for sbottom searches, as was shown in \cite{Gomez:2018zzw,Gomez:2018efz}. This is due to the fact that in  our  scenarios the sbottom squarks are heavy and outside the area covered by the LHC; the same happens with signals involving squarks of the lighter generations.

\begin{figure}
\begin{center}
\includegraphics*[scale=.2]{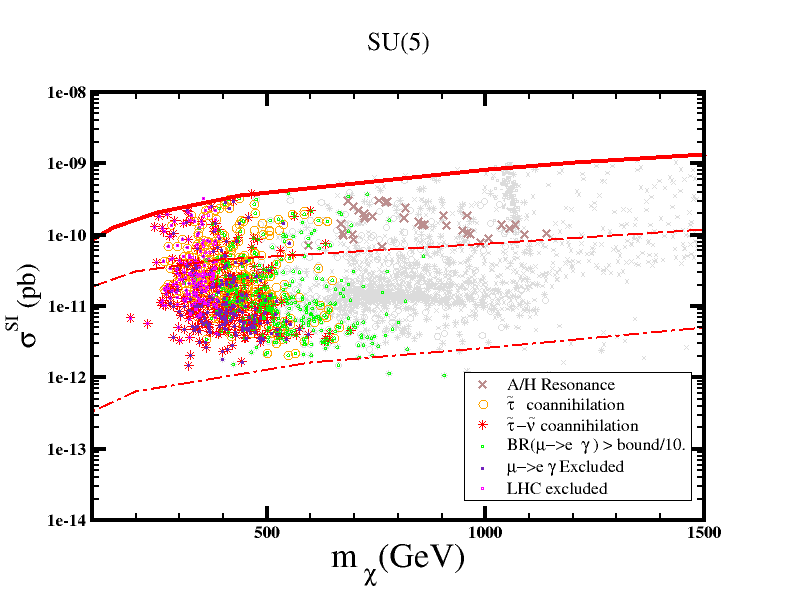}
\includegraphics*[scale=.2]{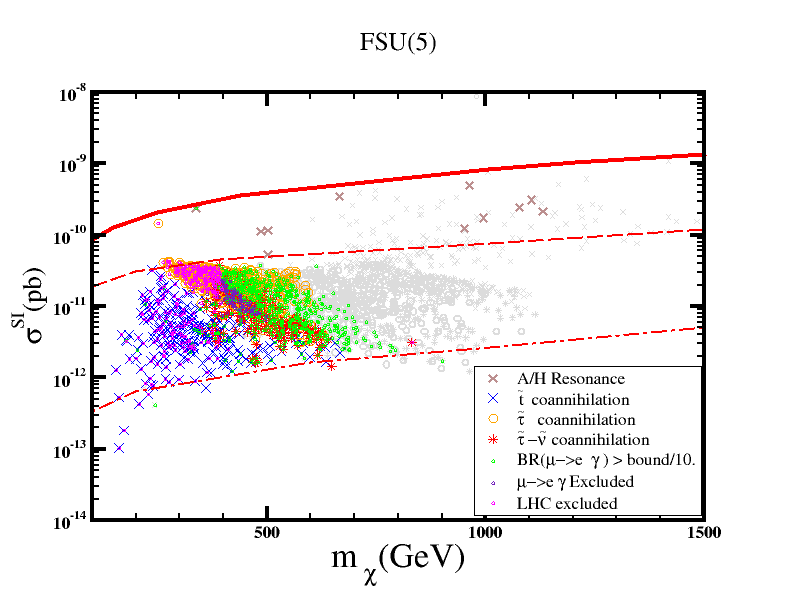}
\includegraphics*[scale=.2]{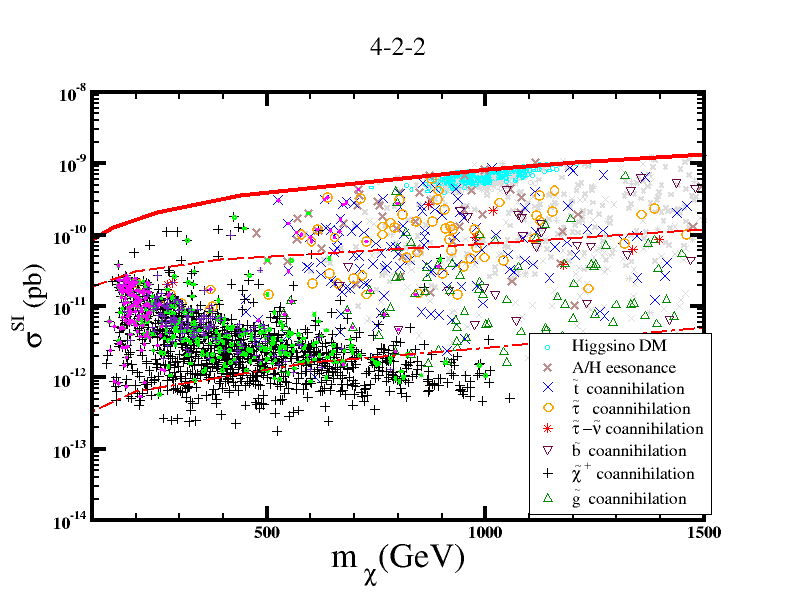}\\
\vspace{-0.5 true cm}
\caption{\it  SI  neutralino-nucleon  cross  section versus $m_\chi$ in SU(5), FSU(5) and the 4-2-2 escenarios. The solid lines corresponds to  the  Xenon-1T  bound~\cite{XENON1T_new}, and the dashed  and  dot-dashed  lines  correspond  to  the  projected  sensitivities  of the  LZ \cite{Akerib:2018dfk}  and DARWIN~\cite{Aalbers:2016jon} experiments.}
\label{fig:si}
\end{center}
\end{figure}

Finally, we display in Fig.~\ref{fig:si} the spin-independent (SI) neutralino-nucleon cross section as a function of the neutralino mass in the different GUT models, and we see that the predictions depend on the unification scenario. We note in particular that the FSU(5) model predicts a lower SI cross section than the SU(5) model, in general, while the 4-2-2 model may yield a relatively large SI cross section even for large neutralino masses $> 1$~TeV. The current bound from the {Xenon-1T} experiment~\cite{XENON1T_new} already excludes many models where the neutralino has a large higgsino component, and the projected sensitivities of the LZ~\cite{Akerib:2018dfk}  and DARWIN~\cite{Aalbers:2016jon} experiments will be able to cover most of the models studied on this work. In particular, only models with $\tilde{\chi}^\pm-\chi$ coannihilations may escape the projected DARWIN sensitivity. Comparing the sensitivities of the LFV, LHC and SI DM searches, we see that the latter are potentially very promising probes of SUSY models. On the other hand, as in~\cite{Gomez:2018efz}, we find in each model that the spin-dependent (SD) neutralino-neutron  cross section is below the projected  limit from the LZ~\cite{Akerib:2018dfk} experiment.

.

\section{Conclusions}

In previous work, we studied the predictions of different unified
theories for DM
and the LHC. We investigated several GUT scenarios,
comparing the areas allowed within different symmetry schemes. We
considered scenarios with gaugino unification,
such as  SU(5) and FSU(5), and models where it can be relaxed, such as 4-2-2 models.   
Among others, we had reached the following conclusions: \\
\hspace*{0.4 cm}
$\bullet $
Models based on SO(10) are very restricted by data, as can be seen in
Refs.~\cite{Ellis:2016tjc, Gomez:2018efz,Gomez:2018zzw}. In contrast, SU(5) models contain several areas of interest for higgsino dark matter,
        resonant annihilations and coannihilations. However, due to
        its multiplet structure, SU(5) models not allow 
        stop-neutralino coannihilations, in contrast to the other
        groups. \\
\hspace*{0.4 cm}
$\bullet $ Flipped SU(5) models can be clearly distinguished from SU(5), and have
        several additional features, including stop-neutralino
        coannihilations. \\
\hspace*{0.4 cm}
$\bullet $
	Models based on 4-2-2 not only give rise to stop-neutralino and sbottom-neutralino coannihilations, they also allow novel DM mechanisms,
 including gluino and chargino coannihilations, as a direct consequence of the distinctive gauge structure.

\hspace*{0.4 cm}

Here we have combined these analyses with the study of LFV,
which turns out to be particularly relevant, using updated LHC
data.  The large mixing for solar and
atmospheric neutrinos implies strong correlations between different
rare decays. Since the limits for $\mu \rightarrow e \gamma$
 are significantly stronger, it made sense to focus mostly on this
 mode and comment on 
$\tau \rightarrow \mu \gamma$ where relevant. We have found the following:

$\bullet $
The three groups have distinctive LFV signatures, making it possible to link specific signatures in rare decays and colliders to the gauge and multiplet structure of the theory.

$\bullet $
The results are naturally sensitive to the scale of the right-handed neutrinos, $M_R$. The see-saw mechanism implies that larger scales are linked to larger couplings and thus larger quantum corrections that violate flavour. For smaller values of $M_R$ the available parameter space is significantly enhanced: indeed, a change of $M_R$ by a factor of 4 is sufficient to exclude or allow a large number of models. 

$\bullet $
In all three groups, coannihilations lead to higher rates for LFV, while resonant annihilations and higgsino dark matter are mostly not affected. Overall, in SU(5) and 4-2-2 it is easier to find annihilation models with good detection prospects both at the LHC and in LFV searches. Higgsino DM models do not predict detectable LFV. Still, it is interesting to note that the LSP composition is different in each scheme, yielding an almost pure Higgsino spectrum in the 4-2-2 model.

$\bullet $
Since the see-saw mechanism introduces LFV only in the LL sector, stau coannihilations with smaller masses and larger left-stau components lead to LFV within the current reach. This is particularly relevant for SU(5) and flipped SU(5), since in 4-2-2 models the left-right splitting of the soft fermion masses makes stau coannihilations more difficult to find. However, the two groups can be clearly distinguished, since SU(5) is more restrictive than flipped SU(5).

$\bullet $
Stop-neutralino coannihilations appear only in flipped SU(5) and 4-2-2 models but, once more, with distinct signatures in each case. In 4-2-2 models the LSP can be heavier, and significantly smaller LFV rates are to be expected.

$\bullet $
The 4-2-2 model also allows chargino and gluino coannihilations with neutralinos, due to the different GUT relations for gaugino masses. Chargino-neutralino coannihilations have good detection prospects for both the LHC and LFV, while gluino coannihilations lead to lower LFV rates.

$\bullet $
There are specific correlations between the sparticle masses, leading
to interesting signatures. In flipped SU(5), gaugino mass
universality results in a proportionality between the gluino and
neutralino masses for most of the models under study (corresponding to
Higgsino DM and resonant annihilations). Larger masses have good LFV
detection prospects, even when they are out of the LHC reach. This is
also true for stop-neutralino coannihilations, as well as for models 
with compressed spectra, such as stau coannihilations.

$\bullet $
In 4-2-2 models, a proportionality relation is found only in
chargino-neutralino coannihilations, again due to the GUT relation when the chargino is
mostly a Wino. 
This class of models provides good prospects for both LFV and the LHC,
 while in other scenarios LFV is significant only for neutralino masses below
 500 GeV. 
This is an additional feature that enables detailed tests of neutralino-chargino coannihilations versus alternative possibilities.

$\bullet$ The experimental advances in direct LSP DM detection are already reaching the sensitivity needed to provide a verdict on many models, specially on the SU(5) and 4-2-2 GUTs. Also, the projected sensitivities of the LZ and DARWIN experiments will provide probes of models that are complementary to LFV searches, even models that cannot be explored at the LHC.\\ 

Overall, our results indicate that LFV is a powerful tool that complements LHC and DM searches, and provides valuable information that can help identify optimal modes for future LHC searches. Moreover, not only does it distinguish clearly between various GUTs
via the observability of different channels, but it can also provide significant insight into the respective sparticle spectra and neutrino mass parameters.

\vspace{1 true cm}

\section*{Acknowledgments}

 The work of J.E. was supported by the UK STFC Grant ST/P000258/1 and by the Estonian Research Council via a Mobilitas Pluss grant. The research of M.E.G. was supported by the Spanish MINECO, under grant FPA2017-86380. R. RdA acknowledges partial funding/support from the Elusives ITN (Marie Sk\l{}odowska-Curie grant agreement No 674896) and the  
``SOM Sabor y origen de la Materia" (FPA 2017-85985-P). Q.S. acknowledges support by the US DOE grant No. DE-SC0013880.


\end{document}